\documentclass[useAMS,usenatbib]{mn2e}

\usepackage{epsfig}
\usepackage{enumerate}
\usepackage{pdflscape}
\usepackage{rotating}

\newcommand{\dd}{deg$^2$ }
\newcommand{\units}{{erg s$^{-1}$ cm$^{-2}$} }
\title[Distant galaxy clusters in the XMM-LSS survey]{Distant galaxy clusters
  in the XMM Large Scale Structure survey}
\author[J. P. Willis et al.]{J. P. Willis$^{1}$\thanks{E-mail: jwillis@uvic.ca (JPW)}, N. Clerc$^{2,3}$, M. N. Bremer$^4$, M. Pierre$^2$, C. Adami$^5$, O. Ilbert$^5$ \newauthor B. Maughan$^4$, S. Maurogordato$^6$, F. Pacaud$^7$, I. Valtchanov$^8$, L. Chiappetti$^9$, \newauthor K. Thanjavur$^{10}$ S. Gwyn$^{11}$, E. R. Stanway$^{4,12}$, C. Winkworth$^4$\\
  $^{1}$Department of Physics and Astronomy, University of Victoria, 3800 Finnerty Road, Victoria, BC, Canada\\
  $^{2}$Service d'Astrophysique, Bât. 709, CEA Saclay, 91191 Gif sur
  Yvette Cedex, France\\
  $^3$Max-Planck Institut fuer Extraterrestrische Physik, Giessenbachstrasse 1, 85748 Garching, Germany\\
  $^4$H. H. Wills Physics Laboratory, University of Bristol, Tyndall
  Avenue, BS8 1TL, UK\\$^5$LAM, OAMP, Universit\'{e} Aix-Marseille \&
  CNRS, P\^{o}le de l'\'{E}toile, Site de Ch\^{a}teau Gombert, \\38
  rue Fr\'{e}d\'{e}ric Joliot-Curie, 13388, Marseille 13 Cedex,
  France\\$^6$Universit\'{e} de Nice Sophia-Antipolis, CNRS,
  Observatoire de la C\^{o}te d'Azur, UMR 6202 Cassiop\'{e}e, BP 4229,
  06304 Nice Cedex 4, France\\$^7$Argelander Institute for Astronomy,
  Bonn University, 53121 Bonn, Germany\\$^8${European Space Astronomy
    Centre, Herschel Science Centre, ESA, 28691 Villanueva de la
    Ca\~nada, Spain}\\$^9$INAF-IASF Milano, Via E. Bassini 15, I-20133
  Milano, Italy\\$^{10}$Canada France Hawaii Telescope Corporation, HI
  96743, USA\\$^{11}$Canadian Astronomy Data Centre, Herzberg
  Institute of Astrophysics, 5071 West Saanich Road, Victoria, British
  Columbia, V9E 2E7, Canada\\$^{12}$Department of Physics, University
  of Warwick, Coventry, CV4 7AL, UK }
\begin{document}

\date{Accepted 4th December 2012. Received 13th November 2012; in original form 24th May 2012}

\pagerange{\pageref{firstpage}--\pageref{lastpage}} \pubyear{2011}

\maketitle

\label{firstpage}

\begin{abstract}
  Distant galaxy clusters provide important tests of the growth of
  large scale structure in addition to highlighting the process of
  galaxy evolution in a consistently defined environment at large look
  back time. We present a sample of 22 distant ($z>0.8$) galaxy
  clusters and cluster candidates selected from the 9 \dd footprint of
  the overlapping X-ray Multi Mirror (XMM) Large Scale Structure
  (LSS), CFHTLS Wide and Spitzer SWIRE surveys. Clusters are selected
  as extended X-ray sources with an accompanying overdensity of
  galaxies displaying optical to mid-infrared photometry consistent
  with $z>0.8$. Nine clusters have confirmed spectroscopic redshifts
  in the interval $0.8<z<1.2$, four of which are presented here for
  the first time.  A further 11 candidate clusters have between 8 and
  10 band photometric redshifts in the interval $0.8<z<2.2$, while the
  remaining two candidates do not have information in sufficient
  wavebands to generate a reliable photometric redshift. All of the
  candidate clusters reported in this paper are presented for the
  first time. Those confirmed and candidate clusters with available
  near infrared photometry display evidence for a red sequence galaxy
  population, determined either individually or via a stacking
  analysis, whose colour is consistent with the expectation of an old,
  coeval stellar population observed at the cluster redshift.  We
  further note that the sample displays a large range of red fraction
  values indicating that the clusters may be at different stages of
  red sequence assembly. We compare the observed X-ray emission to the
  flux expected from a suite of model clusters and find that the
  sample displays an effective mass limit $M_{200} \sim 1 \times
  10^{14} \rm M_\odot$ with all clusters displaying masses consistent
  with $M_{200}<5 \times 10^{14} \rm M_\odot$. This XMM distant
  cluster study represents a complete sample of X-ray selected $z>0.8$
  clusters. We discuss the importance of this sample to investigate
  the abundance of high redshift clusters and to provide a relatively
  unbiased view of distant cluster galaxy populations.
\end{abstract}


\section{Introduction}

Observations of galaxy clusters provide crucial insight into the
development of structure in the Universe, from the growth of clusters
themselves, to the evolution of their member galaxies. Furthermore,
cluster studies yield important constraints on cosmological models
through tests of the growth of structure (e.g. \citealt{vik09};
\citealt{pierre11}). The greatest constraining power on cosmological
parameters and on the co-evolution of galaxies and clusters requires
large, well-controlled samples of clusters out to $z > 1$.
To date, a number of techniques have been successfully applied in
order to generate such samples of clusters at $z<1$. These include
approaches based upon detecting galaxy overdensities (in a combination
of apparent colour and sky position; e.g. \citealt{post96,glad05}),
extended X-ray sources (e.g. \citealt{gioia90,bohr00,bur07,mehrtens11}) and
the Sunyaev-Zel'dovich decrement observed toward the Cosmic Microwave
Background (e.g. \citealt{menanteau10,reichardt12}).

Although the systematic estimation of the cluster number density above
$z > 1$ is an important issue the search for high redshift clusters is
made difficult by the faintness of the cluster signal, e.g. small
galaxy over-density in optical and near-infrared (NIR) imaging, weak
X-ray emission whose extent is difficult to assess, etc.  However,
should X-ray imaging observations of sufficient depth and spatial
resolution be executed, the detection of high redshift clusters via
spatially-extended emission is advantageous as it provides clear
evidence of hot gas confined within a gravitational potential well.
Care must be taken though to assess the extent to which faint X-ray
active galactic nuclei (AGN) may mimic or modify the significance and
spatial extent of cluster emission and its spectral form
(e.g. \citealt{branch07,pierre12}).

Systematic searches for high redshift galaxy clusters in X-rays are
currently being conducted via dedicated XMM surveys (XDCP
\citealt{fass11}; XMM-LSS \citealt{pierre07,bremer06,andreon05}, this
work; XCS \citealt{romer01,stanford06}). These surveys employ
quantitative algorithms to identify extended sources and are
characterised by accurate selection functions and a clearly defined
relationship between cluster observables (such as X-ray luminosity)
and total cluster mass \citep{reichert2011}.  Galaxy clusters at $z>1$
may also be detected by the extension of successful ``red sequence''
searches to near-- and mid--infrared (MIR) wavebands in order to
detected the redshifted emission from distant galaxies
(e.g. \citealt{muzzin09}). An associated technique employed by the
IRAC Shallow Cluster Survey \citep{eisenhardt2008} identifes clusters
to $z<1.5$ as stellar mass overdensities in multi-band photometric
redshift slices.  Each MIR imaging technique has proven very
successful at identifying large numbers of candidate and confirmed
clusters. 

Galaxy clusters at $z<1$ have been employed extensively to study their
member galaxy populations and indicate that they are composed of
uniformly old stellar populations where the bulk of their stars formed
at $z=3$ or greater \citep{jaffe11}.  In addition, low redshift
clusters display strong population trends such as the
morphology--density relation (e.g. \citealt{dressler97}).  Such
relations reflect the dominance of bright, red, bulge-dominated
galaxies in cluster cores. Observing clusters at high redshift
provides an opportunity to approach the epoch when the progenitors of
low-redshift galaxies where assembled.  Indeed impressive direct
evidence is emerging of both red sequence truncation \citep{rudnick12}
and merger driven galaxy assembly \citep{lotz12} in a MIR selected
(yet X-ray detected) cluster at $z=1.6$. The evolution of the
brightest cluster galaxies (BCGs) in distant clusters presents a more
complex picture: \citet{lidman2012} report that the stellar mass of
BCGs in 160 clusters spanning $0.03<z<1.63$ grow steadily with
decreasing redshift in a manner consistent with a semi-analytic model.
In potential contrast to such evolution in high redshift cluster
galaxies, \citet{stott10} report that the stellar mass contained in
the BCGs of a sample of 20, $0.8<z<1.5$ clusters
has changed little between the epoch of observation and the present
day.  Such trends observed in heterogeneously assembled samples will
be better understood by performing similarly extensive analyses upon a
complete sample of high redshift clusters selected employing a single
method (e.g. \citealt{fass11}).

The XMM-LSS survey is well placed to contribute to this investigation:
covering 11 \dd with X-ray imaging to a depth of $\sim 1 \times
10^{-14}$ \units for extended sources in the [0.5-2]keV waveband and
accompanied by optical and MIR photometry. The XMM-LSS project has
previously demonstrated the ability to identify clusters to $z=1.2$
(\citealt{bremer06}) and has published $z<1$ cluster number counts
selected according to a clear, quantitative selection function
(\citealt{pacaud07}).

This paper presents the XMM-LSS distant cluster sample. These are
defined to be extended X-ray sources at $z>0.8$ and consist of
spectroscopically confirmed clusters together with a number of
candidate clusters supported by a detailed photometric redshift
analysis.  The distant cluster sample has been identified from the
full sample of extended X-ray sources within a 9 \dd\ sub-area of the
XMM-LSS survey and in this sense it represents a complete sample of
X-ray selected distant clusters.  In particular, the methods employed
to determine whether a given extended X-ray source is or is not a
distant cluster are selected to minimise any potential bias such as
the presence of a strong red sequence.  In this sense the sample
should be as complete as possible and should provide an unbiased
perspective of the galaxy populations in distant X-ray clusters.

The structure of the paper is as follows. Section 2 describes the
construction of the distant cluster sample (containing both
spectroscopically confirmed and candidate systems) and the
multi-wavelength data sets used to define it in addition to presenting
images of the sources in the sample. Section 3 describes the
approaches taken to explore which of the candidate systems have clear
photometric evidence for being genuine distant clusters. Section 4
discusses the results of applying these approaches and explores the
diversity of properties shown by the likely high redshift systems.

This paper employs a Friedmann-Robertson-Walker cosmological model
described by the parameters $\Omega_M=0.3$, $\Omega_\Lambda=0.7$,
$H_0=70 \rm \, kms^{-1}Mpc^{-1}$.  An angular scale of 1\arcmin\
corresponds to a transverse physical scale of 480 kpc, 508 kpc and 502
kpc at redshifts $z=1$, 1.5 and 2 respectively. All photometry is
quoted in the AB system.

\section{The XMM-LSS survey sample}

The XMM-LSS survey currently covers approximately 11 \dd and is
described in \citet{chia12} and Clerc et al. (2013). Galaxy clusters
are detected as extended X-ray sources and are classified as either C1
or C2 on the basis of their surface brightness characteristics
\citep{pacaud06}.  The effective flux limit is $\sim 1\times
10^{-14}$\units for extended sources.

Visual inspection of the X-ray images of individual systems along with
their optical and NIR images confirm that the C1 class represents an
uncontaminated sample of extended X-ray sources (mainly clusters but
with some detections of X-ray halos in very low redshift galaxies,
\citealt{pacaud06}). The C2 class displays a contamination rate of
30-50\%, with the main sources of contamination being misclassified
point sources and artefacts on the X-ray image. These contaminants are
typically removed by visual inspection of the X-ray image prior to
further analysis.

The 11 \dd XMM-LSS sample contains 50 C1 and 60 C2 sources, of which
44 C1 and 27 C2 sources have confirmed redshifts from optical
spectroscopy. The redshift distribution of confirmed sources ranges
over $0.05<z<1.22$.  The lower spectroscopic confirmation rate for the
C2 sources arises due to a) the lower priority placed on follow-up of
such sources compared to C1 sources and b) the increased difficulty
of following up fainter, lower quality detections. 

The analysis used to generate the XMM-LSS distant cluster sample is
based upon a 9 \dd subregion of the XMM-LSS field.  This region
represents the common footprints of the XMM-LSS, Canada France Hawaii
Telescope Legacy Survey (CFHTLS) W1 and Spitzer SWIRE 3.6 and
4.5$\micron$ surveys and contains 88 C1$+$C2 sources (of which 55 have
spectroscopic redshifts).  NIR imaging drawn from a variety of
sources (i.e. the UKIDSS and WIRDS surveys in addition to individual
CFHT/WIRCam and VLT/HAWKI images) exists for many of the
spectroscopically confirmed and candidate distant systems, with the
available bands and depth varying on a source-by source basis. The
principal data sets and their processing are described below.


\subsection{X-ray photometry}

The characterisation and measurement of extended sources in the
XMM-LSS survey is described in detail in \cite{pacaud06} and
\cite{adami11} and we summarise the important features of the analysis
here.  Sources are detected above a specified pixel threshold by
appliying the {\tt SExtractor} routine to a multi-scale wavelet
reconstruction of the XMM science image.  Individual sources are
characterised as either extended or point-like on the basis of the
likelihood values of appropriate models applied to each source.
Taking the example of an extended source, nearby point-sources
identified individually by the detection algorithm are masked when
performing the source extended fit using the {\tt SExtractor}
segmentation map \citep{pacaud06}. Therefore the extent likelihood
used for C1/C2 classification is almost free from contamination by
bright point sources. Faint point-sources contaminating the extended
source emission and not deblended by the algorithm may be present,
particularly in regions close to the cluster centre. Accounting for
them is challenging given the faintness of our objects and their
compactness relative to XMM point spread function.

We compute X-ray fluxes for the extended sources associated with
spectroscopically confirmed and candidate clusters in the [0.5-2] keV
band employing the procedure outlined in \citet{adami11}. The method
applies a curve-of-growth analysis to the X-ray count rate data which
confers the advantage of being free of any profile assumptions applied
to the extended X-ray source.  We estimate that the application of a
finite size aperture used for these measurements recovers 80-90\% of
the total count rate of the cluster and we note that this bias
is lower than the statistical error of our measurements.  The flux
measurement step allows a further check for additional blended
emission.  Point sources lying close to but off the cluster central
region are either identified by the detection algorithm or flagged
visually. In both cases their contribution to the total flux is
excluded from the extraction region, and the missing area is accounted
for by assuming a circularly symmetric flux profile.  The fluxes were
obtained assuming a fixed conversion factor of $9 \times 10^{−13} \rm
(erg^{}s^{-1}cm^{-2})/(cnts~s^{-1})$. This value was calculated using
{\tt Xspec} from an APEC emission model with the following parameters:
$z=1$, $T=4$ keV, $N_H = 2.6 \times 10^{20} \rm cm^{−2}$, $Ab = 0.3$
(note that the conversion factor changes by less than 5\% for $z=1.5$
and $T=4$ keV). Bolometric luminosities were calculated with {\tt
  Xspec} employing the measured fluxes, the cluster redshift (quoted
in Table \ref{tab_clus}) and the fitted cluster temperature
(\citealt{bremer06,pacaud07}) or assuming $T=4$ keV if no temperature
was available.  Flux and luminosity values for each cluster are listed
in Table \ref{tab_clus}.

\subsection{Multi-wavelength photometric data}
\label{sec_nir}

As noted above, most of the data used in this analysis are taken from
existing large area surveys which cover part or all of the XMM-LSS
region.  Descriptions of the CFHTLS and Spitzer SWIRE data used in the
paper are described in Gwyn (2012) and Chiappetti et al. (2012)
respectively.  Large area NIR survey data provided $J$, $H$ and/or
$K$-band imaging for a number of the C1 and C2 $z>0.8$
candidates. These data came from either the UKIDSS Deep Extragalactic
Survey \citep[DXS DR8; see][]{lawrence07} or the WIRCam Deep Survey
\citep[WIRDS;][]{bielby10}.  Typical depths ($5\sigma$) for these data
are $J=23.4,K=22.9$ (DXS) and $J=24.7, H=24.7, Ks=24.7$ (WIRDS).  NIR
imaging data for the remaining sources in the distant cluster sample
were obtained with principal investigator programs using VLT/HAWKI and
CFHT/WIRCam and are described below. Details of the available
multi-wavelength photometry for each $z>0.8$ system, both candidate
and confirmed, are given in Table \ref{tab_clus}.

%
%

\begin{landscape}
\begin{table}
  \caption{Distant clusters in the XMM-LSS survey. These sources
    represent all C1$+$C2 sources (Section 2) within the approximately
    9 \dd XMM-LSS/CHFTLS/SWIRE footprint that cannot be unambiguously
    associated with a $z<0.8$ cluster or galaxy. Spectroscopically
    confirmed distant clusters are displayed in the upper section of
    the table and have their redshift values indicated. Each is also
    described by an official XLSSC identifier. Candidate distant
    clusters are displayed in the middle section of the table and
    their photometric redshift values (where computed) are indicated
    with accompanying uncertainty. Unknown sources are displayed in
    the lower section of the table and do not have a redshift
    value. Each cluster is further identified using a sequential
    number (column 1) to simplify their discussion in this paper. In
    subseqent publications that refer to these sources, we recommend
    using either the cluster names or the XLSSC numbers which are the
    only IAU validated labelings. The procedures used to compute the
    X-ray flux and luminosity measures are described in Section 2.1.}
\label{tab_clus}
\centering
\begin{tabular}{clccccccclll}
\hline
Number & Cluster name & XLSSC & Class & R.A. & Dec. & redshift & Flux [0.5-2]keV &  $L_X,bol$ & wavebands & Reference for \\
&&&&(J2000)&(J2000)&& $\times 10^{-14}$\units & $\times 10^{44}$erg s$^{-1}$&&spectroscopic redshift \\
\hline
01 &XLSS\hspace{2.5mm}  J022400.4-032529   & 032 & C2 & 36.002 & -3.424 & 0.803    & $1.56\pm0.43$ & $1.16\pm0.32$ & $grz3.6\micron4.5\micron$       & Section 2.4 \\
02 &XLSS\hspace{2.5mm}  J022233.8-045803   & 066 & C2 & 35.641 & -4.968 & 0.833    & $1.13\pm0.27$ & $0.92\pm0.22$ & $ugrizK3.6\micron4.5\micron$    & Section 2.4 \\
03 &XLSSU J021832.0-050105                 & 064 & C2 & 34.633 & -5.106 & 0.875    & $1.48\pm0.12$ & $1.36\pm0.11$ & $ugriz3.6\micron4.5\micron$     & Adami et al. (2011) \\
04 &XLSSU J021524.1-034332                 & 067 & C1 & 33.850 & -3.726 & 1.003    & $6.26\pm0.64$ & $7.85\pm0.80$ & $ugrizYJK_s3.6\micron4.5\micron$  & Section 2.4 \\
05 &XLSS\hspace{2.5mm} J022253.6-032828    & 048 & C1 & 33.850 & -3.726 & 1.005    & $1.66\pm0.36$ & $2.11\pm0.46$ & $ugriz3.6\micron4.5\micron$     & Pacaud et al. (2007) \\
06 &XLSSU J021458.6-033020                 & 068 & C2 & 33.745 & -3.506 & 1.032    & $0.67\pm0.17$ & $0.90\pm0.23$ & $grzYJK_s3.6\micron$              & Section 2.4 \\
07 &XLSS\hspace{2.5mm}  J022404.1-041330   & 029 & C1 & 36.017 & -4.225 & 1.050    & $3.56\pm0.33$ & $5.02\pm0.46$ & $ugrizJK3.6\micron4.5\micron$   & Pierre et al. (2007)\\
08 &XLSS\hspace{2.5mm}  J022709.2-041800   & 005 & C1 & 36.788 & -4.300 & 1.053    & $1.03\pm0.15$ & $1.47\pm0.22$ & $ugrizJHK_s3.6\micron4.5\micron$  & Valtchanov et al. (2004)\\
09 &XLSS\hspace{2.5mm}  J022303.3-043621   & 046 & C2 & 35.764 & -4.606 & 1.213    & $0.60\pm0.17$ & $1.20\pm0.34$ & $ugrizJK_s3.6\micron4.5\micron$   & Bremer et al. (2006)\\
\hline
10 &XLSS\hspace{2.5mm}  J021721.4-050855   &     & C2 & 34.340 & -5.149 & $0.65^{+0.12}_{-0.12}$ & $0.39\pm0.07$ & $0.18\pm0.03$ & $ugrizJK_s3.6\micron4.5\micron$   & \\
11 &XLSSU J022411.5-045327                 &     & C2 & 36.048 & -4.891 & $0.70^{+0.18}_{-0.16}$ & $0.58\pm0.21$ & $0.32\pm0.11$ & $ugrizJHK_s3.6\micron4.5\micron$  & \\
12 &XLSSU J021547.7-045027                 &     & C1 & 33.948 & -4.842 & $0.96^{+0.19}_{-0.21}$ & $1.12\pm0.22$ & $1.27\pm0.25$ & $ugrizJK_s3.6\micron4.5\micron$   & \\
13 &XLSSU J021859.5-034608                 &     & C2 & 34.748 & -3.769 & $0.99^{+0.21}_{-0.19}$ & $1.19\pm0.25$ & $1.46\pm0.31$ & $ugrizJK3.6\micron4.5\micron$   & \\
14 &XLSS\hspace{2.5mm}  J022059.0-043921   &     & C2 & 35.245 & -4.656 & $1.11^{+0.29}_{-0.26}$ & $0.86\pm0.16$ & $1.37\pm0.25$ & $ugrizJK_s3.6\micron4.5\micron$   & \\
15 &XLSS\hspace{2.5mm}  J022252.3-041647   &     & C2 & 35.718 & -4.280 & $1.12^{+0.18}_{-0.17}$ & $0.55\pm0.13$ & $0.89\pm0.22$ & $ugrizJK_s3.6\micron4.5\micron$   & \\
16 &XLSSU J021712.1-041059                 &     & C2 & 34.300 & -4.183 & $1.48^{+0.25}_{-0.10}$ & $0.58\pm0.19$ & $1.87\pm0.63$ & $ugrizYJK_s3.6\micron4.5\micron$  & \\
17 &XLSSU J021700.3-034747                 &     & C2 & 34.252 & -3.797 & $1.54^{+0.30}_{-0.31}$ & $0.49\pm0.18$ & $1.73\pm0.64$ & $ugrizYJK_s3.6\micron4.5\micron$  & \\
18 &XLSSU J022005.5-050824                 &     & C2 & 35.024 & -5.141 & $1.65^{+0.25}_{-0.26}$ & $0.81\pm0.21$ & $3.40\pm0.86$ & $ugrizYJK_s3.6\micron4.5\micron$  & \\
19 &XLSS\hspace{2.5mm}  J022812.1-043845   &     & C2 & 37.050 & -4.646 & $1.67^{+0.20}_{-0.20}$ & $0.57\pm0.17$ & $2.44\pm0.75$ & $ugrizYJK_s$                      & \\
20 &XLSS\hspace{2.5mm}  J022418.7-043959   &     & C2 & 36.078 & -4.666 & $1.67^{+0.20}_{-0.20}$ & $0.38\pm0.24$ & $1.65\pm1.03$ & $ugrizJHK_s3.6\micron4.5\micron$  & \\
21 &XLSSU J021744.1-034536                 &     & C1 & 34.433 & -3.760 & $1.91^{+0.19}_{-0.21}$ & $1.08\pm0.27$ & $6.48\pm1.64$ & $ugrizYJK_s3.6\micron4.5\micron$  & \\
22 &XLSS\hspace{2.5mm}  J022554.5-045058   &     & C2 & 36.477 & -4.849 & $2.24^{+0.26}_{-0.24}$ & $0.20\pm0.06$ & $1.79\pm0.51$ & $ugrizYJK_s3.6\micron4.5\micron$  & \\
23 &XLSS\hspace{2.5mm}  J022227.9-051554   &     & C2 & 35.616 & -5.265 & N/A       & $0.70\pm0.13$ & N/A           & $ugriz3.6\micron4.5\micron$     & \\
24 &XLSSU J022200.8-040636                 &     & C2 & 35.503 & -4.112 & N/A       & $0.31\pm0.10$ & N/A           & $ugrizK3.6\micron4.5\micron$    & \\
\hline
25 &XLSS\hspace{2.5mm} J022351.3-041840    &     & C2 & 35.964 & -4.312 & N/A      & $0.99\pm0.14$ & N/A           & $ugrizJK3.6\micron4.5\micron$   & \\
26 &XLSS\hspace{2.5mm} J022127.6-043258    &     & C2 & 35.365 & -4.550 & N/A      & $0.30\pm0.13$ & N/A           & $ugriz3.6\micron4.5\micron$     & \\
27 &XLSS\hspace{2.5mm} J021944.4-043943    &     & C2 & 34.935 & -4.662 & N/A      & $0.31\pm0.15$ & N/A           & $ugriz3.6\micron4.5\micron$     & \\
28 &XLSS\hspace{2.5mm} J022111.6-034223    &     & C2 & 35.298 & -3.707 & N/A      & $0.82\pm0.15$ & N/A           & $ugriz3.6\micron4.5\micron$     & \\
29 &XLSS\hspace{2.5mm} J022712.8-044632    &     & C2 & 36.803 & -4.775 & N/A      & $1.47\pm0.23$ & N/A           & $ugrizJHK_s3.6\micron4.5\micron$  & \\
30 &XLSS\hspace{2.5mm} J022339.3-035918    &     & C2 & 35.913 & -3.989 & N/A      & $0.56\pm0.19$ & N/A           & $ugrizK3.6\micron4.5\micron$    & \\
31 &XLSS\hspace{2.5mm} J022034.3-040544    &     & C2 & 35.142 & -4.095 & N/A      & $0.26\pm0.07$ & N/A           & $ugrizJK3.6\micron4.5\micron$   & \\
32 &XLSS\hspace{2.5mm} J022803.9-051740    &     & C2 & 37.016 & -5.295 & N/A      & $1.32\pm0.30$ & N/A           & $ugriz3.6\micron4.5\micron$     & \\
\hline
\end{tabular}
\end{table}

\end{landscape}

Eight of the C1 and C2 X-ray sources that were candidate or confirmed
$z>0.8$ clusters were observed using HAWK-I on the VLT through ESO
programme 084.A-0740(A). The HAWK-I camera consists of four HAWAII 2RG
$2048\times2048$ pixel detectors. The four detectors image an area of
$7\farcm5 \times 7\farcm5$ with a pixel scale of $0\farcs106$
pixel$^{-1}$ (the cross shaped gap between the four detectors is
15\arcsec\ wide). Each candidate was observed using the $YJK_S$
filters with exposure times of 1800 seconds, 2530 seconds, and 3600
seconds respectively and a suitable offset was applied to each set of
observations to place the measured X-ray centroid of each cluster
candidate in the centre of a single detector.  Data were obtained in
service mode during October 2009 to January 2010.

Reduction was carried our using standard procedures from version 1.4.2
of the ESO pipeline within the {\tt esorex} environment. The pipeline
routines corrected the data for the presence of bad pixels, dark
current, flat field variations, two-stage sky subtraction with object
masking, distortion correction and co-addition with pixel rejection.

Photometric calibration was a two-step process. Firstly, standard star
observations were used to place all four detectors on a common
photometric scale. Then the $J$ and $K_s$-band images were placed on
the 2MASS scale by matching stars with $J<15.8$ and $K_s<14.3$ to their
counterparts in the HAWK-I images. Between 3 and 20 such stars were
present in each image and this procedure resulted in zeropoints
accurate to better than 0.1 mags in all cases. The official HAWK-I
$Y$-band zeropoint is yet to be made available. In its absence, the
procedures used in \citet{hickey10} were used. Using the HAWK-I
$J$-band photometry and the CFHTLS $z$-band, a pseudo $Y$-band
magnitude was constructed from their flux average for each
source. Using only those sources with $z-J\sim 0$, the HAWK-I $Y$ zero
point was adjusted to match the pseudo-$Y$ photometry. The typical
sensitivity of these data are $Y=24.3,J=24.0,K_s=23.5$ ($5\sigma$)
within a 3\arcsec\ diameter aperture.

A further 4 candidate C2 distant clusters were observed with
CFHT/WIRCam in December-January 2011-12. The WIRCam camera consists of
four HAWAII2-RG detectors, each containing $2048 \times 2048$
pixels. The four detectors image an area of $20\arcmin \times
20\arcmin$ with a pixel scale of $0\farcs3$ pixel$^{-1}$ (the cross
shaped gap between the four detectors is 45\arcsec\ wide). Each
candidate was observed using the $JK_s$ filters with exposure times of
5428 to 8614 seconds, and 3675 seconds respectively. A suitable
offset was applied to each set of observations to place the measured
X-ray centroid of each cluster candidate in the centre of a single
detector. Photometric calibration followed the method outlined above
and generated typical sensitivity values $J=23.0,K_s=22.5$ ($5\sigma$)
within a 3\arcsec\ diameter aperture.

\subsection{Catalogue construction}
\label{sec_cat}

Optical and NIR imaging data for each X-ray source were placed on a
common pixel scale using the {\tt swarp v2.17.1} software package
\citep{bertin02}.  Source extraction and photometry were then
performed using {\tt SExtractor v2.5.0} \citep{bertin96} employed in
dual-image mode using the $K_S$- or $K$-band image as the detection
image in each case\footnote{Note that for brevity in the following
  text we refer to $K_s$ photometry for all relevant sources as this
  is the $K$-band filter predominantly used and as the colour term
  between the UKIDSS $K$ filter and the $K_s$ filters used for
  CFHT/WIRCam and VLT/HAWK-I is small for the $z>0.8$ galaxies of
  interest in this paper.}.  Photometry was computed within an
aperture based upon the Kron (1980) radius.  The image quality of the
HAWK-I and WIRCam data are well matched to that of the CFHTLS optical
data, with typical stellar FWHM values of $0\farcs6-0\farcs7$.

Finally, we match the optical-NIR catalogues for each field to sources
detected at 3.6\micron\ and 4.5\micron\ in the SWIRE
catalogue. Sources are matched with a $1\farcs5$ tolerance and Spitzer
photometry is quoted within a 2\arcsec\ radius circular aperture with
an additive offset applied to correct to a pseudo-total aperture.
We note that this approach may introduce a small additive offset
between the matched optical-NIR and the Sptizer photometry and we
attempt to determine and correct for any such zero-point offsets
between wavebands within the photometric redshift analysis.

\subsection{Spectroscopic observations}

Nine galaxy clusters presented in this distant cluster sample have
confirmed spectroscopic redshifts. Five have been published previously
and appropriate references are provided in Table
\ref{tab_clus}. Spectroscopic data for four of the clusters are
presented here for the first time and we decribe the observation,
reduction and analysis of the data below.

Details of the spectroscopic observations obtained for clusters 01,
02, 04 and 06 are presented in Table \ref{tab_spec_conf}. Slit targets
for each cluster consisted of galaxies located within the X-ray
emitting isophotes with photometric redshifts consistent with being
cluster members. Further slits were placed upon moderately bright
stars and galaxies in order to provide identifiable reference
spectra. The data were processed using standard techniques in the {\tt
  IRAF}\footnote{IRAF is distributed by the National Optical Astronomy
  Observatories, which are operated by the Association of Universities
  for Research in Astronomy, Inc., under cooperative agreement with
  the National Science Foundation.}  environment which included
procedures to extract one dimensional spectra, apply a wavelength
solution based upon reference HeAr spectra (or sky features in the
case of GMOS-S spectra) and to correct for varying spectrograph
response using observations of a spectrophotometric standard star. The
spectral resolution of each data set was determined by measuring the
full-width at half-maximum (FWHM) of unresolved arc lines.
\begin{table*}
  \caption{Details of the spectroscopic observations performed on four clusters in the sample.}
\label{tab_spec_conf}
\begin{tabular}{cccccccccc}
\hline
Cluster & Observing date & Telescope/Spectrograph & Grism$+$Filter & Wavelength & Spectral & Exposure \\
&&&& coverage (\AA) & resolution (\AA) & time (s) \\
\hline
01  & November 2003 & VLT/FORS2 & 600RI$+$GG435 & 5000-8500 & 7 & 3600 \\
02  & November 2003 & VLT/FORS2 & 600RI$+$GG435 & 5000-8500 & 7 & 3600 \\
04 & November/December 2006 & Gemini/GMOS-S & R400$+$OG515 & 6000-10000 & 8 & 17500 \\
06 & November 2010 & VLT/FORS2 & 300I$+$OG590 & 6000-10000 & 20 & 16200 \\
\hline
\end{tabular}
\end{table*}

Reduced spectra were inspected visually to provide an initial estimate
of galaxy redshifts based upon the identification of prominent
features. Individual spectra were then cross--correlated with a
representative early--type galaxy template (e.g. \citealt{kinney96})
employing routines based upon \citet{tonry79}.  Individual cluster
members were selected by identifying visually an initial cluster
redshift $z_{peak}$ in the redshift histogram for each field. Cluster
galaxies were then selected to display $z_{peak}\pm 0.03$ and
individual redshift values are displayed in Table
\ref{tab_spec_values}. The cluster redshifts presented in Table
\ref{tab_clus} were then computed as the mean redshift of all members
located within 1\arcmin\ of the cluster X-ray centroid.
\begin{table}
  \caption{Spectroscopic redshifts for individual galaxies in each cluster.}
\label{tab_spec_values}
\begin{tabular}{cccc}
\hline
ID & R.A. (J2000) & Dec. (J2000) & redshift \\
\hline
01 & 35.9829 & -3.3481 & 0.799\\
01 & 35.9870 & -3.3455 & 0.801\\
01 & 36.0002 & -3.3564 & 0.800\\
01 & 36.0026 & -3.4265 & 0.799\\
01 & 36.0042 & -3.4272 & 0.807\\
01 & 36.0003 & -3.4263 & 0.803\\
01 & 36.0373 & -3.3946 & 0.801\\
01 & 36.0108 & -3.4389 & 0.803\\
\hline
02 & 35.6420 & -4.9655 & 0.832\\
02 & 35.6430 & -4.9689 & 0.825\\
02 & 35.6397 & -4.9720 & 0.842\\
02 & 35.6401 & -4.9587 & 0.832\\
02 & 35.6414 & -4.9775 & 0.843\\
02 & 35.6345 & -4.9761 & 0.833\\
02 & 35.6832 & -5.0105 & 0.831\\
02 & 35.6714 & -5.0079 & 0.822\\
02 & 35.6121 & -4.9991 & 0.861\\
02 & 35.6657 & -5.0137 & 0.844\\
02 & 35.6114 & -4.9987 & 0.862\\
02 & 35.6413 & -4.9652 & 0.834\\
02 & 35.6824 & -5.0102 & 0.832\\
\hline
04 & 33.8869 & -3.7395 & 1.000\\
04 & 33.8826 & -3.7344 & 1.008\\
04 & 33.8801 & -3.7197 & 1.004\\
04 & 33.8771 & -3.7355 & 0.999\\
04 & 33.8593 & -3.7478 & 1.010\\
04 & 33.8571 & -3.7290 & 1.002\\
04 & 33.8535 & -3.7269 & 1.006\\
04 & 33.8494 & -3.7268 & 1.003\\
04 & 33.8461 & -3.7275 & 0.989\\
04 & 33.8435 & -3.7203 & 0.999\\
04 & 33.8412 & -3.7321 & 1.002\\
04 & 33.8352 & -3.7298 & 1.011\\
04 & 33.8231 & -3.7332 & 1.000\\
04 & 33.8070 & -3.7649 & 0.999\\
\hline
06 & 33.7371 & -3.5019 & 1.033\\
06 & 33.7406 & -3.4999 & 1.032\\
06 & 33.7439 & -3.5019 & 1.029\\
06 & 33.7460 & -3.5062 & 1.033\\
\hline
\end{tabular}
\end{table}

\section{Identifying distant cluster candidates}

The aim of this paper is to present a complete sample of extended
X-ray sources with redshifts $z>0.8$.  The motivation for generating a
complete sample essentially centres upon the assertion that, only by
generating a complete sample of X-ray selected $z>0.8$ clusters can
one a) compare the abundance of distant clusters to a cosmological
model prediction in quantitative manner and b) discuss the range of
galaxy properties exhibited by the member population in a relatively
unbiased way.  Constructing a complete sample ultimately requires
providing a robust explanation of the nature of every extended X-ray
source in the sample area.  Though it is the aim of the XMM-LSS survey
to confirm spectroscopically all extended C1 and C2 sources, this
remains an observationally challenging prospect at this stage.  The
methodology of this paper is therefore to employ the available
photometric data to direct the detailed follow-up of individual
clusters (e.g. deep NIR imaging and spectroscopic observations)
towards extended sources that show compelling evidence for being a
bona-fide cluster above some redshift threshold.

The threshold of $z>0.8$ applied in this paper was defined in response
to largely practical considerations, e.g.  given the depth of
available optical data, experience indicated that all clusters at
$z<0.8$ could be recognised with little ambiguity.  A threshold of
$z>0.8$ (as opposed to a larger value) generated a final sample
containing approximately 20 distant clusters which was deemed large
enough that evolution of cluster galaxy properties such as red
sequence colour and population mix could be traced within the sample
rather than via reference to other cluster samples selected possibly
using alternative techniques.  Finally a threshold of $z>0.8$
corresponds to a look back time of approximately 7 Gyr and accords
with the redshift definition applied to distant clusters in the
literature.

We summarise below the steps employed to evaluate the robustness (or
otherwise) of the evidence pointing to a high redshift identification
for each candidate:

\begin{itemize}

\item Classify visually all C1/C2 sources based on $rz3.6\micron$ and
  $3.6\micron4.5\micron$ images. Identify potentially distant systems
  and reject misclassified point sources.

\item Compute the surface density of galaxies along the line of sight
  to each source that satisfy photometric selection criteria
  appropriate for identifying $z>0.8$ galaxies. Compare to the visual
  classification results as a check.

\item Compute the colour of galaxy overdensities along the line of
  sight to each source and compare to the expected colour of model
  galaxies as a test of the distant cluster hypothesis.

\item Compute photometric redshifts for galaxies along the line of
  sight to distant cluster candidates with available multi-wavelength
  photometry. Identify photometric redshift peaks spatially associated
  with the location of the extended X-ray source.

\item Obtain where possible spectroscopic observations of distant
  cluster candidates with $z_{phot}>0.8$ with the aim of confirming at
  least 3 concordant redshifts within the sky area giving rise to the
  extended X-ray emission.

\end{itemize}


\subsection{Visual classification}

The first step is straightforward and is carried out on all C1 and C2
sources in the 9 \dd XMM-LSS/CFHTLS/SWIRE field. The CFHTLS and SWIRE
fields containing each C1 and C2 X-ray source were inspected visually,
in order to identify any obvious clustering at or close to the X-ray
position. Images in individual bands and pseudo-true colour images
($r/z/3.6\micron$ and $r/3.6\micron/4.5\micron$) were used in this
process. This step was performed by up to six people and the final
classification was subject to the decision of two moderators.

Visual inspection is relatively rapid to perform and provides useful
information on the broad nature of each X-ray source e.g. bright,
clustered galaxies consistent with a low redshift cluster, a
misclassified point source, or a significant, extended X-ray source
with at best a grouping of faint, red galaxies consistent with being a
high-redshift cluster. However, visual classification is likely to
provide only an imprecise estimate of the redshift of each system. In
practice, though all 88 C1$+$C2 sources were inspected visually, the
classification efforts were focussed on the 33/88 C1$+$C2 sources
lacking a secure spectroscopic identification.

\begin{figure}
\centering
\psfig{figure=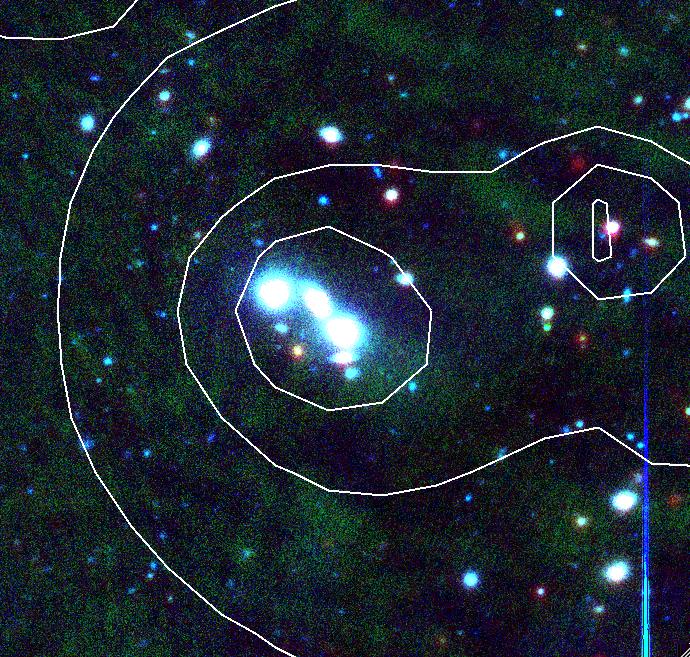,width=3.0in,angle=0.0}
\caption{An example image of one of the 12 extended X-ray sources
  unambiguously associated with a $z<0.8$ cluster on the basis of
  visual inspection. The image is composed of $rz3.6\micron$ data and
  is 2\arcmin\ on a side with standard astronomical
  orientation.  The white contours indicate X-ray emission. }
\label{fig_ex_loz}
\end{figure}

Of these 33 sources, 12 displayed convincing evidence for being a
cluster at a redshift sufficiently below $z=0.8$ to be a secure
classification (Figure \ref{fig_ex_loz} provides an image of a typical
system).  The remaining sources displayed either a) weak or absent
optical emission with an identifiable clustering of
3.6\micron\ sources, b) a clear case of a misclassified point source,
or c) a weak, yet potentially extended X-ray source with no
identifiable overdensity of galaxies at 3.6\micron.  Of these classes
of objects, those in group (a) are retained as candidate $z>0.8$
clusters (we note at this stage that the visual classification may
include sources at $z<0.8$ but the philosophy at this stage is that
any cut avoid being conservative).  Sources in group (c) could either
represent false detections resulting from the X-ray pipeline or
potentially very distant X-ray clusters with little evidence of galaxy
clustering at any wavelength. In either case, the prospects for
spectroscopic confirmation of such systems are exceptionally poor and
\---\ as they represent at best 1 or 2 systems out of a sample of 88
\---\ are not retained as candidate distant clusters at this stage.

\begin{figure}
\centering
\psfig{figure=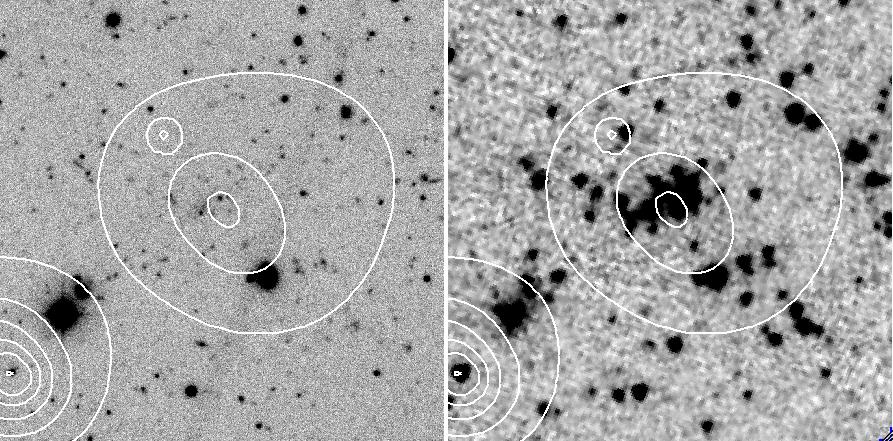,width=3.3in,angle=0.0}
\caption{A comparison of the appearance of the confirmed $z=1.2$
  cluster XLSSC 046 in $r$-band (left panel) and the
  3.6\micron\ waveband (right panel). Each image is approximately
  2\arcmin\ on a side with standard astronomical orientation.  The
  white contours indicate X-ray emission. }
\label{fig_bremer}
\end{figure}

This simple sifting can work to at least $z=1.2$ as the highest
redshift XMM-LSS cluster XLSSC 046 \citep{bremer06} is
straightforwardly selected in this way with an obvious compact
overdensity of $3.6\micron$ galaxies at the X-ray position with no
bright optical counterparts (see Figure \ref{fig_bremer}). However, at
high redshifts whether a cluster is straightforwardly identifiable
depends on several factors affecting its surface density contrast
against the background and foreground galaxies. If the
foreground/background density in the $z=1.2$ cluster field had been
higher and the spatial distribution of red cluster galaxies less
compact (spread over a 1\arcmin\ radius region rather than the
15\arcsec\ radius of the detected overdensity), the system would have
been harder to discern in this way.

Two exceptions to this process are clusters 06 and 19. Cluster 06
lies at the very edge of the SWIRE footprint and has data available at
3.6\micron\ but not 4.5\micron. Cluster candidate 19 lies just
outside the SWIRE footprint yet had previously been flagged as a high
redshift candidate on the basis of the extended X-ray image and faint,
$i$-band detection of the candidate BCG. We include it in the
following discussion and in Table \ref{tab_clus} yet do not include it
in the number of systems quoted in the $z>0.8$ area limited sample.

\begin{figure*}
\centering
\psfig{figure=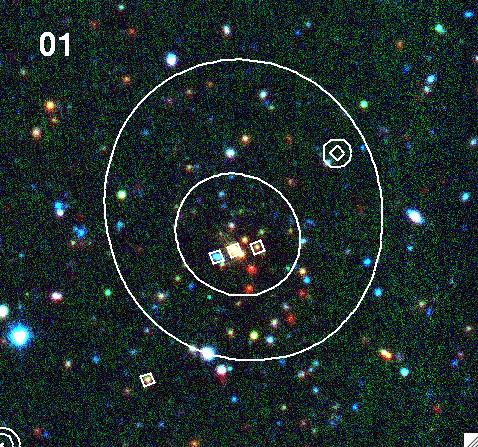,width=2.2in,angle=0.0}
\psfig{figure=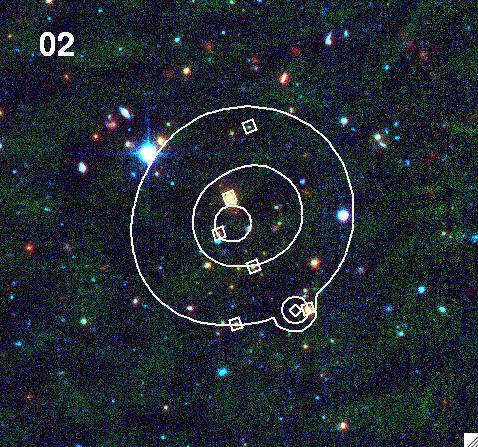,width=2.2in,angle=0.0}
\psfig{figure=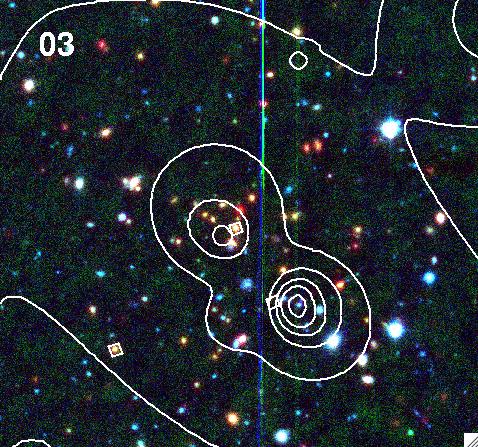,width=2.2in,angle=0.0}
\psfig{figure=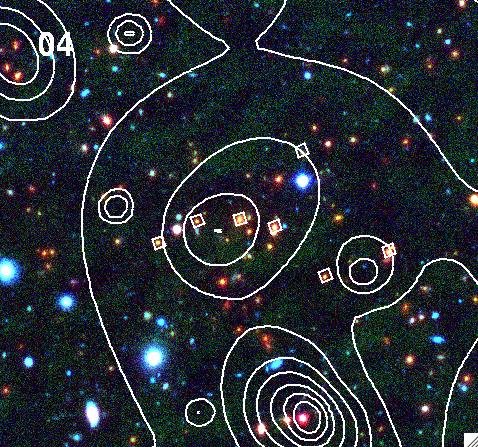,width=2.2in,angle=0.0}
\psfig{figure=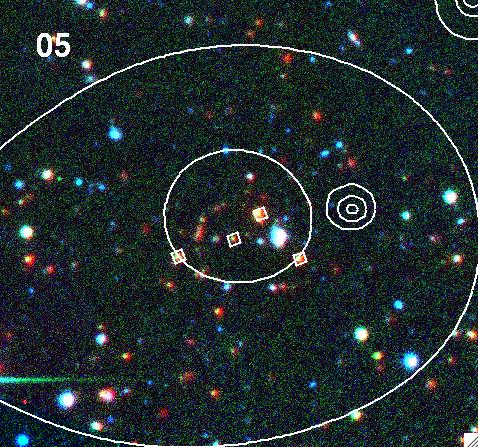,width=2.2in,angle=0.0}
\psfig{figure=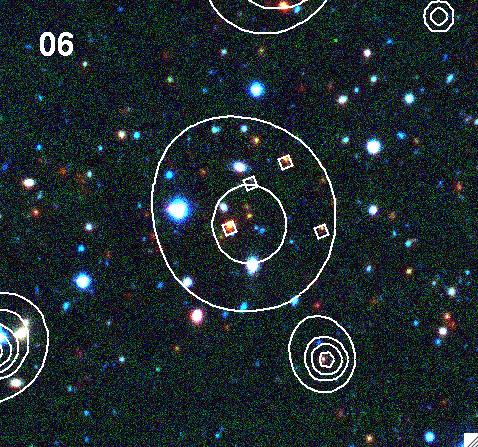,width=2.2in,angle=0.0}
\psfig{figure=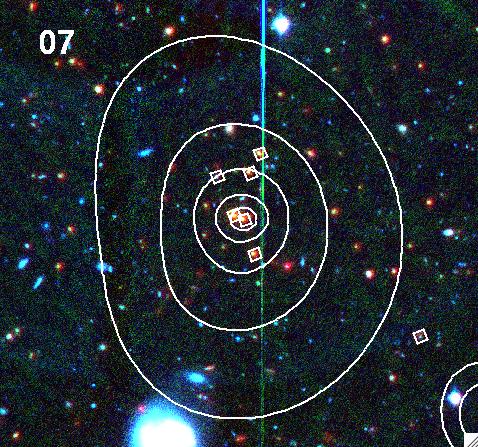,width=2.2in,angle=0.0}
\psfig{figure=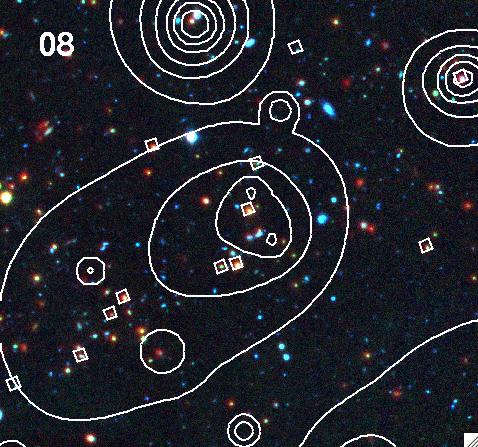,width=2.2in,angle=0.0}
\psfig{figure=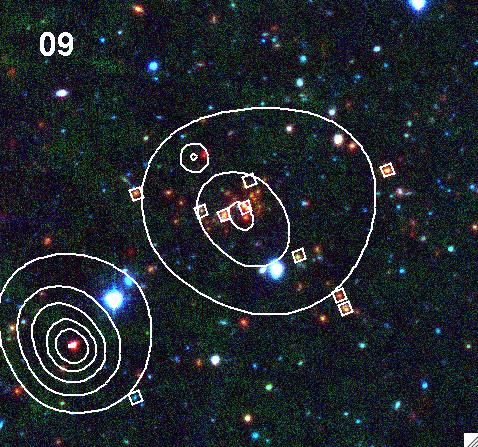,width=2.2in,angle=0.0}
\caption{Colour images of spectroscopically confirmed clusters (see
  Table \ref{tab_clus}). Images are composed of $rz3.6\micron$
  data. The white squares indicate spectroscopically confirmed
  galaxies at the cluster redshift. The white contours indicate X-ray
  emission. Images are 2\arcmin\ on a side with standard
  astronomical orientation.}

\label{fig_spec_rzI1}
\end{figure*}

\begin{figure*}
\centering
\psfig{figure=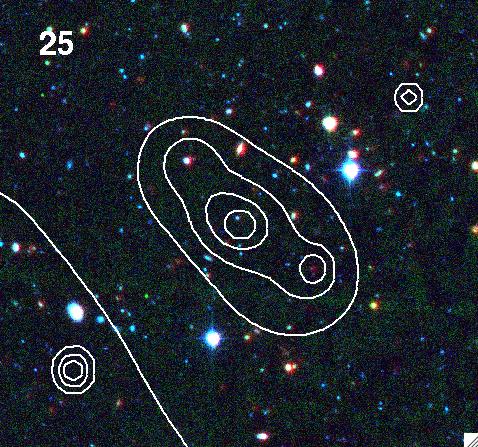,width=2.2in,angle=0.0}
\psfig{figure=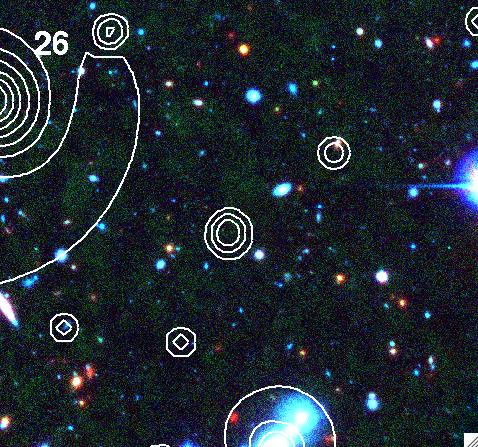,width=2.2in,angle=0.0}
\psfig{figure=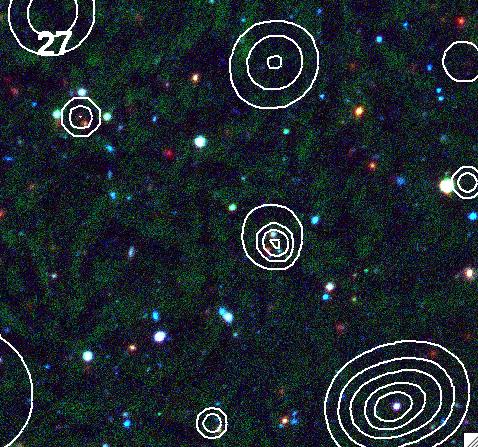,width=2.2in,angle=0.0}
\psfig{figure=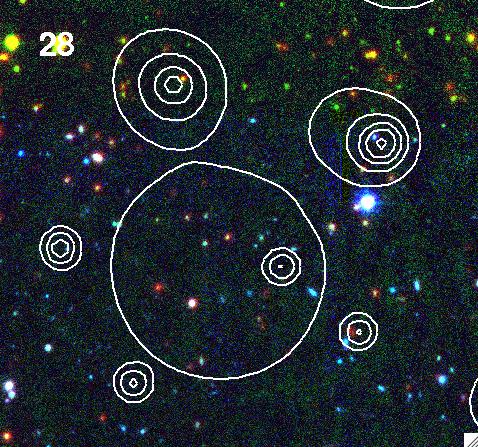,width=2.2in,angle=0.0}
\psfig{figure=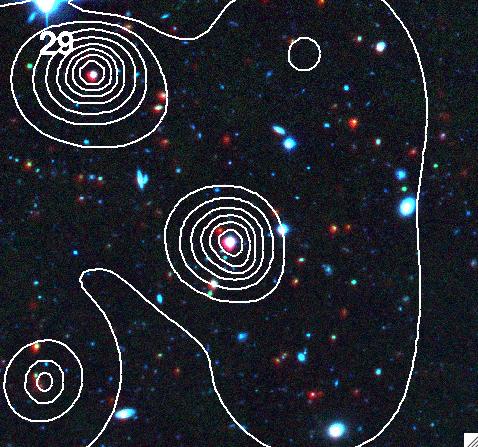,width=2.2in,angle=0.0}
\psfig{figure=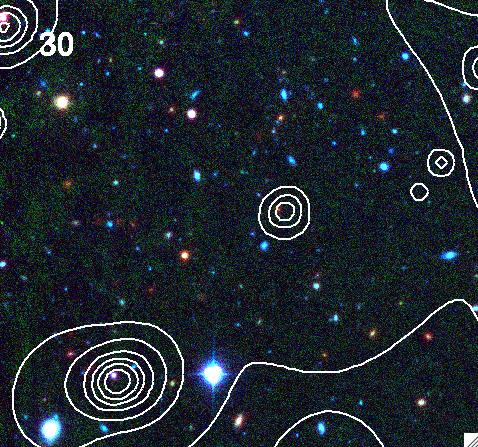,width=2.2in,angle=0.0}
\psfig{figure=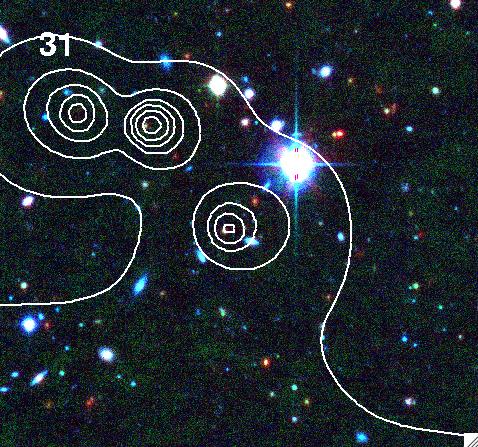,width=2.2in,angle=0.0}
\psfig{figure=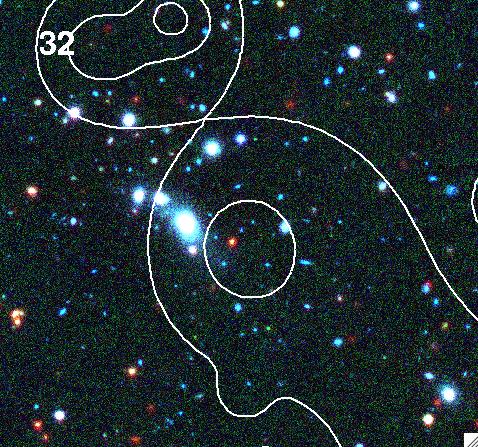,width=2.2in,angle=0.0}
\caption{Colour images of marginal or unknown extended X-ray sources
  (see Table \ref{tab_clus}). Images are composed of $rz3.6\micron$
  data.  The white contours indicate X-ray emission. Images are
  2\arcmin\ on a side with standard astronomical
  orientation.}
\label{fig_dud_rzI1}
\end{figure*}

The visual classification supported by the existing spectroscopic
observations populates each cluster class as follows:
spectroscopically confirmed or candidate clusters at $z<0.8$, 57
(``$z<0.8$ clusters'' hereafter); spectroscopically confirmed clusters
at $z>0.8$, 9 (see Figure \ref{fig_spec_rzI1}; ``confirmed clusters''
hereafter); candidate clusters at $z>0.8$, 14 (see Figure
\ref{fig_cand_rzI1}; ``candidate clusters'' hereafter); point-like or
marginal sources with no clear identification, 8, (see Figure
\ref{fig_dud_rzI1}; ``unknown extended sources'' hereafter).

\begin{figure*}
\centering
\psfig{figure=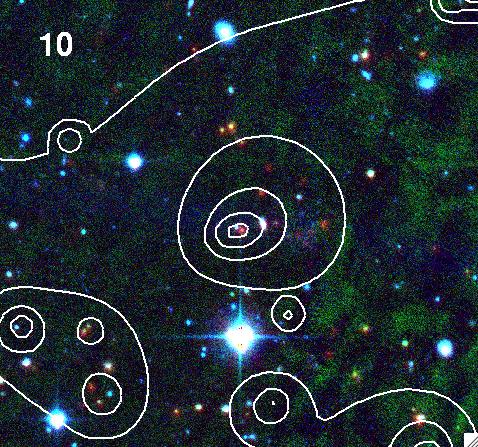,width=2.2in,angle=0.0}
\psfig{figure=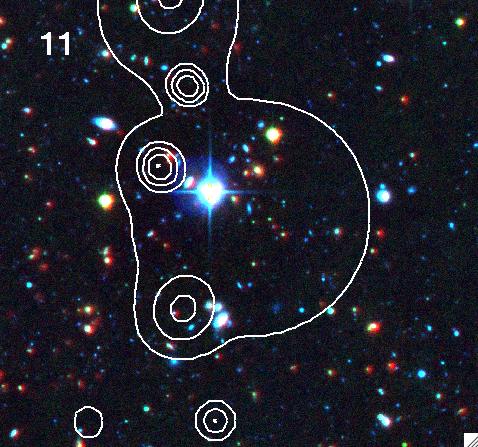,width=2.2in,angle=0.0}
\psfig{figure=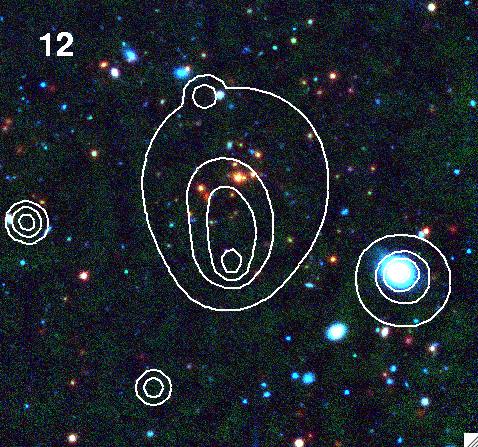,width=2.2in,angle=0.0}
\psfig{figure=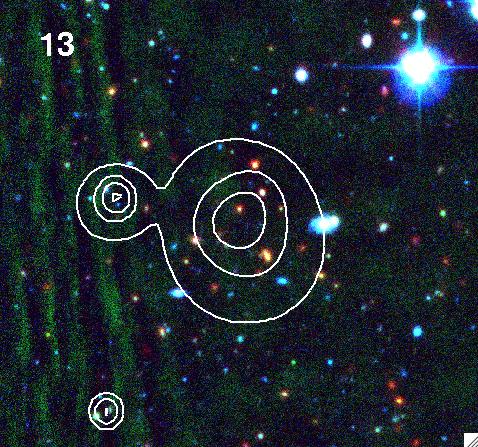,width=2.2in,angle=0.0}
\psfig{figure=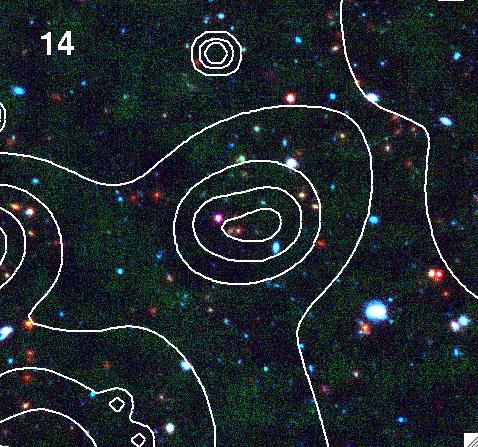,width=2.2in,angle=0.0}
\psfig{figure=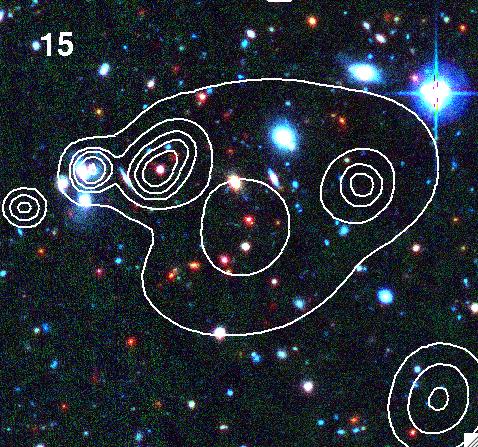,width=2.2in,angle=0.0}
\psfig{figure=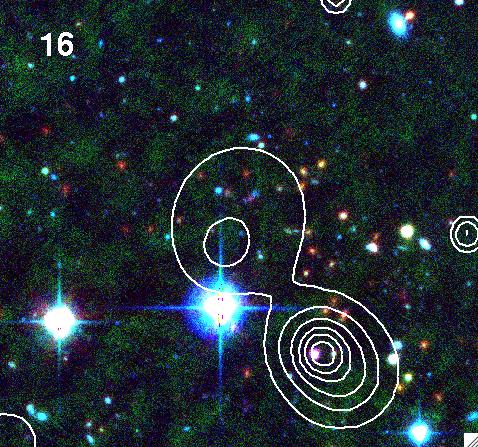,width=2.2in,angle=0.0}
\psfig{figure=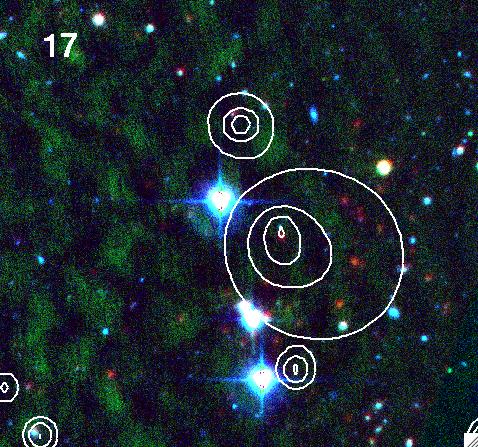,width=2.2in,angle=0.0}
\psfig{figure=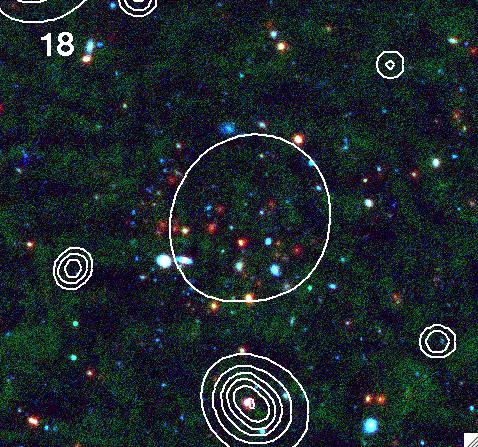,width=2.2in,angle=0.0}
\psfig{figure=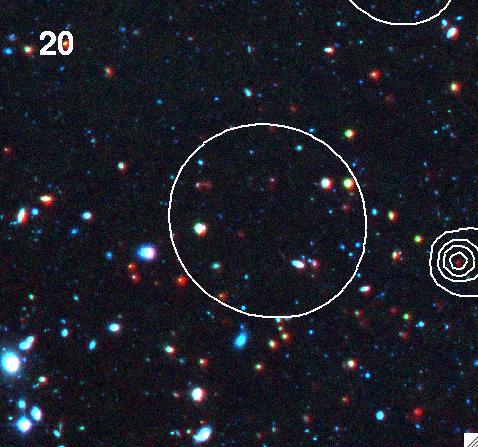,width=2.2in,angle=0.0}
\psfig{figure=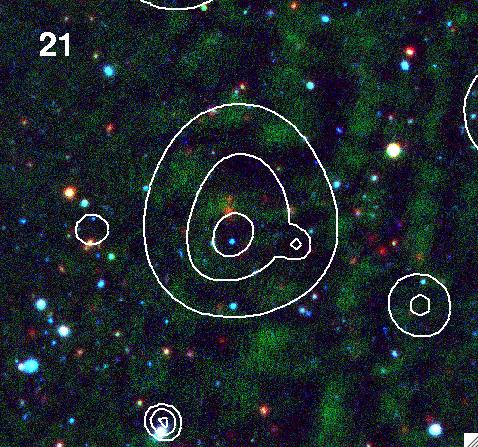,width=2.2in,angle=0.0}
\psfig{figure=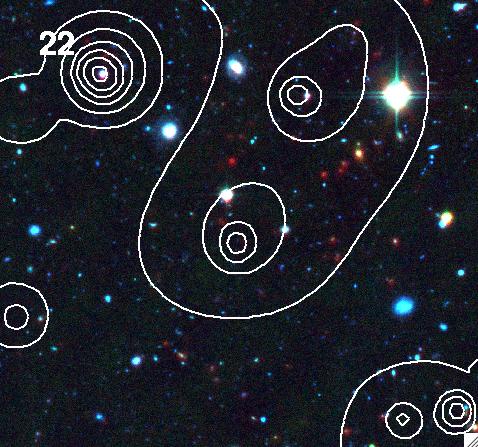,width=2.2in,angle=0.0}
\caption{Colour images of candidate clusters (see Table
  \ref{tab_clus}). Images are composed of $rz3.6\micron$ data.  The
  white contours indicate X-ray emission. Images are approximately
  2\arcmin\ on a side with standard astronomical orientation.}

\label{fig_cand_rzI1}
\end{figure*}

\begin{figure*}
\centering
\psfig{figure=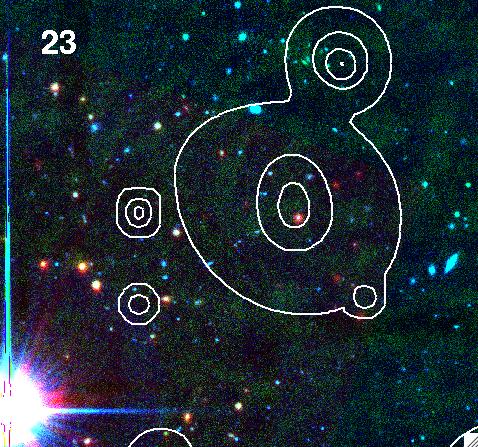,width=2.2in,angle=0.0}
\psfig{figure=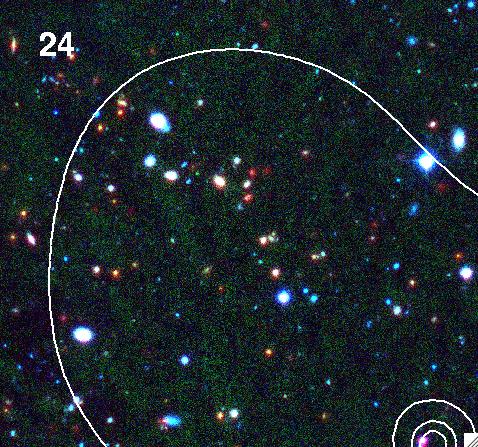,width=2.2in,angle=0.0}
\contcaption{Colour images of candidate clusters (see Table
  \ref{tab_clus}). Images are composed of $rz[3.6\micron]$ data.  The
  white contours indicate X-ray emission. Images are
  2\arcmin\ on a side with standard astronomical
  orientation.}

\end{figure*}

\subsection{The surface density of faint, red galaxies}

Following the visual inspection, the colours of galaxies in all
fields, whether or not identified as clustered from initial
inspection, are examined in order to identify any clustering in both
position and colour. These colours, given data of suitable depth,
can provide an indication of the redshift of the system.

\begin{figure}
\centering
\psfig{figure=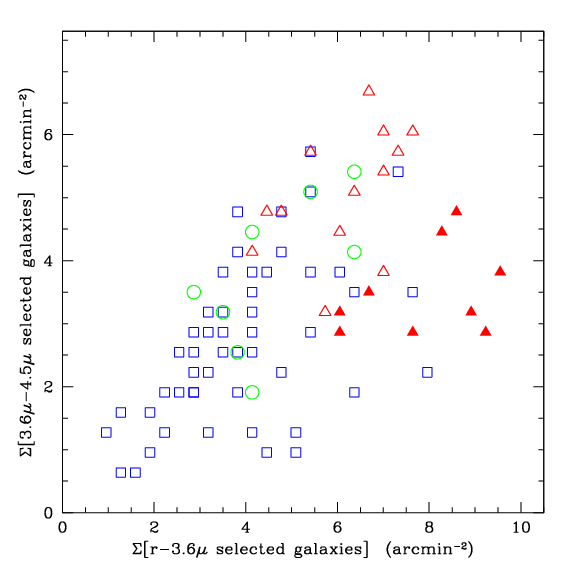,width=3.0in,angle=0.0}
\caption{The surface density of galaxies selected according to the
  photometric criteria described in the text and located within
  1\arcmin\ of the X-ray centroid. Sources are classified as $z<0.8$ (blue
  squares), $z>0.8$ spectroscopically confirmed clusters (solid red
  triangles), $z>0.8$ candidate clusters (open red triangles) and
  unknown sources (green circles).}
\label{fig_apc}
\end{figure}

The photometric analysis employs the $r-3.6\micron$ and
$3.6\micron-4.5\micron$ colours derived from the CFHTLS and
SPITZER/IRAC data. These colours have been proven to select for
distant cluster galaxy populations \citep[see][]{pap08,muzzin08}.
With data of sufficient depth, the use of the $3.6\micron-4.5\micron$
colour is particularly powerful as it is a strong function of redshift
between $0.5<z<1.5$ for galaxies with a wide range of star formation
histories and the scatter in colour between histories is relatively
small \citep[see][]{pap08}.  For all clusters, the surface density of
galaxies within 1\arcmin\ of the X-ray position with $r-3.6\micron>3$
and (independently) $r>22$ and $3.6\micron-4.5\micron$$>-0.1$ was
computed.  For 3.6\micron\ sources that are undetected in $r$
(approximately $r>25.9$) the computed colour is a lower limit on the
true colour.  These values were compared to those computed for the
$z<0.8$ clusters and 1000 randomly placed apertures over the survey
area. Figure \ref{fig_apc} displays these values for all C1 and C2
clusters and cluster candidates in the 9 \dd\ area. As expected, the
spectroscopically confirmed distant clusters are in the top-right of
the distribution, along with many of the candidates. This is evidence
that at least a subset of the candidates are at redshifts similar to,
or even higher than, those of the spectroscopically confirmed
clusters, with the candidate fields having a higher surface density of
sources with the reddest $3.6\micron-4.5\micron$ colours than the
spectroscopically confirmed $z>0.8$ clusters.

The separation of cluster classes indicated in Figure \ref{fig_apc} is
supported by a two-sided Kolmogorov-Smirnov (KS) analysis of the
surface density values (in this case using $r-3.6\micron$ colours). A
KS test between the $z<0.8$ cluster sample and the 1000 random
apertures result in a probability that the two samples are drawn from
the same population of 0.74, which is sensible given the photometric
criteria are designed to be sensitive to $z>0.8$ galaxies. The
corresponding KS probabilities between the $z>0.8$ sample (confirmed
and candidates) and the $z<0.8$ clusters, the random apertures and the
unknown sources are $5 \times 10^{-4}$, $5 \times 10^{-5}$ and $4
\times 10^{-2}$ respectively \---\ thus confirming the trend observed
in Figure \ref{fig_apc}.

However, some overlap remains between the visually-assigned
classes. For the low- versus high-redshift sources this is partly due
to the fact that photometric uncertainties will result in low redshift
galaxies exceeding the applied colour cut and vice versa \---\ this
blurring is expected to be most evident for clusters close to the
effective redshift cut-off implied by the photometric criteria.  In
addition, while it is unlikely that any of the 57 $z<0.8$ clusters are
at $z>0.8$ we do note that two of the $z>0.8$ candidate clusters
ultimately result in photometric redshift estimates slightly less than
$z=0.8$.  In general though, when considering the $z>0.8$ candidate
clusters and the unknown extended sources the surface density analysis
indicates that the optical-MIR data cannot provide an unambiguous
assessment of these systems.  Put another way, there is no
straightforward threshold that can be applied to the surface density
of optical-MIR selected galaxies along the line of sight to the
extended X-ray source sample that will generate a complete sample of
candidate distant cluster candidates with low contamination. We
address this point further in Section \ref{discuss}.

\subsection{Identifying cluster red sequences}
\label{sec_rzI1}

\begin{figure}
\centering
\psfig{figure=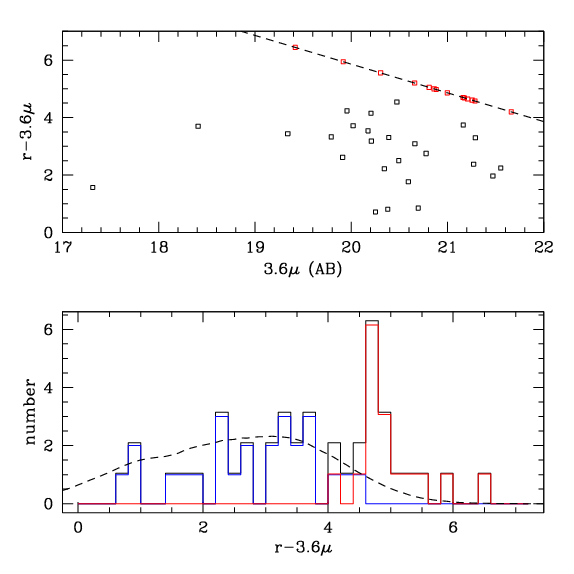,width=3.5in,angle=0.0}
\caption{Optical-MIR photometry for cluster candidate 21. All sources
  within 1\arcmin\ of the X-ray position are indicated. Top panel: red
  points indicate sources undetected in $r$. Bottom panel: the
  histograms indicate sources detected in $r$ (blue), undetected in
  $r$ (red) and total source counts (black). Each histogram is scaled
  to be visible in overlay. The dashed line shows the background
  colour distribution. This is formed by computing the colour
  histogram of all sources more than 1\arcmin\ away from each C1$+$C2
  source in the sample. The histogram amplitude is then scaled by the
  ratio of the areas of the cluster and background areas.}
\label{fig_cmd_col_hist}
\end{figure}

\begin{figure}
\centering
\psfig{figure=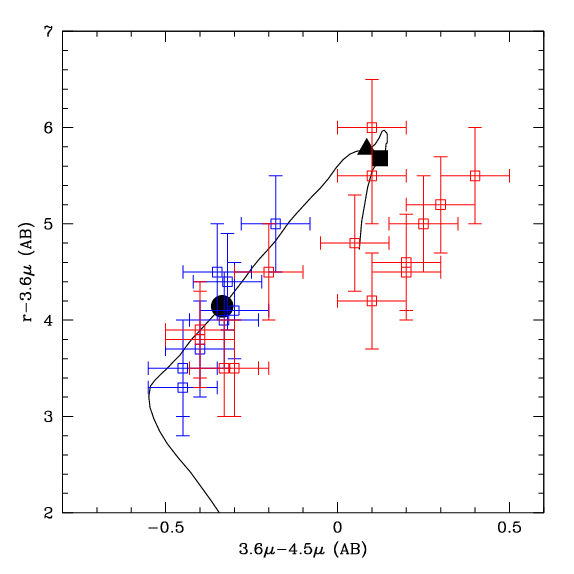,width=3.5in,angle=0.0}
\caption{The distribution of cluster red sequence locations in the
  $r-3.6\micron$ and $3.6\micron-4.5\micron$ plane. Blue points
  indicate spectroscopically confirmed clusters at $z>0.8$ and red
  points indicate candidate clusters at $z>0.8$. Errors are fixed at
  0.1 mag. in $3.6\micron-4.5\micron$ and 0.5 mag. in
  $r-3.6\micron$. The black line indicates the locus followed by a
  passively evolving 1 Gyr solar metallicity burst of star formation
  occuring at $z_f=5$ \citep{bc03}. The colour of this model stellar
  population at redshifts 1, 1.5 and 2 are indicated respectively by
  the black circle, triangle and square.}
\label{fig_hiz_2col}
\end{figure}

In addition to computing the surface density of faint, red galaxies,
colour magnitude diagrams (CMDs) and colour histograms in both
$r-3.6\micron$ and $3.6\micron-4.5\micron$ were created for the
confirmed and candidate $z>0.8$ clusters and compared to the
background colour distribution (e.g. see Figure
\ref{fig_cmd_col_hist}).  Assuming that a cluster contains a
significant number of passively-evolving galaxies, the presence of a
red sequence should confirm a cluster identification and the
characteristics of that sequence should indicate an estimated redshift
for the cluster when compared to stellar population models (Figure
\ref{fig_hiz_2col}). The analysis indicates that the confirmed
clusters, located at $0.8<z<1.2$, display red sequence colours broadly
consistent with the expectation of a simple model of an old, passively
evolving stellar population considered at the confirmed spectroscopic
redshift (see Figure \ref{fig_hiz_2col} for further details).  A
number of the candidate clusters at $z>0.8$ also display red sequences
consistent with the expectation of a high redshift passive stellar
population.  However, it is also clear that a subset of the candidate
clusters do not display red sequence colours consistent with this
simple model of a high redshift stellar population \---\ specifically
they appear to be systematically bluer in $r-3.6\micron$ than the
expected colour for $1<z<2$. This effect can be understood by noting
(as indicated in Figure \ref{fig_cmd_col_hist}) that a number of
sources detected at 3.6\micron\ in the candidate clusters are
undetected in CFHTLS W1 $r$-band data. The indicated $r-3.6\micron$
red sequence colours are therefore lower limits and will underestimate
the true colour.  A further limitation is that the
$3.6\micron-4.5\micron$ red sequence colour is largely degenerate with
redshift at $z \ga 1.3$.  The above steps indicate (often strongly)
that each confirmed and candidate cluster is associated with an
overdensity of galaxies consistent with $z>0.8$.  However, although
the optical and MIR imaging data are effective at determining the
presence of over-densities of high-redshift galaxies they only provide
limited information on the redshift associated with the red sequence
location.

\subsection{Photometric redshift analysis}

Having searched for evidence of red sequences in colour magnitude
diagrams (CMDs), we extend our analysis by removing the assumption
that any distant cluster galaxy population has a significant red
sequence, and allow the possibility that the galaxies associated
with a cluster can have a range of spectral energy distributions
corresponding to a range of star formation histories. 

We use the public code Le Phare \citep{arnouts02,ilbert06} to estimate
the photometric redshifts. Le Phare is based on a standard template
fitting procedure. The templates are redshifted and integrated through
the appropriate transmission curves. The photometric redshifts are
obtained by comparing the modelled fluxes and the observed fluxes with
a $\chi^2$ merit function. We run the code using exactly the same
configuration as used in the COSMOS field \citep{ilbert09}. The set of
templates was generated by \citet{polletta07} for the elliptical and
spiral galaxies. Twelve blue templates generated with \citet{bc03}
were added. Four different dust extinction laws were applied
\citep{prevot84,calzetti00}, and an additional bump at 2175\AA,
depending on the considered template.  Emission lines were added to
the templates using relations between the UV continuum, the star
formation rate and the emission line fluxes \citep{ken98}.  Moreover,
an automatic calibration of the zero-points was performed using a
sample of 650 spectroscopic redshifts within the photometric data area
drawn from the XMM-LSS sample described by \cite{adami11}.  The
calibration is obtained by comparing the observed and modelled fluxes
\citep{ilbert06} at known spectroscopic redshifts.

Having demonstrated that the optical and MIR colours of candidate
$z>0.8$ galaxies provide only a relatively inaccurate estimate of the
redshift of a given confirmed or candidate cluster, the photometric
redshift analysis described here is limited to those confirmed and
candidate clusters with additional NIR photometry in at least the $J$
and $K_s$ bands (see Table \ref{tab_clus} for a list of clusters with
$JK$ photometry).  At $z>0.8$ the $JK$ bands sample the rest frame
optical spectral energy distribution (SED) including the prominent
D4000 feature which provides a strong $JK$ colour signature \---\ and
thus strong constraining power in a photometric redshift analysis
\---\ as a function of redshift for galaxies composed of evolved
stellar populations.  Figure \ref{fig:3panel} displays the photometric
redshift histograms for all clusters with available $JK$ data.  For
clusters with spectroscopic redshifts, data are plotted for all
galaxies within 30\arcsec\ of the X-ray source that are brighter than
the $K_s$-band completeness limit of each data set (see Section
\ref{sec_nir}). For clusters with photometric redshifts (see below)
$z_{phot}<1.2$ and $z_{phot}>1.2$, data are plotted for all galaxies
within 1\arcmin\ and 30\arcsec\ respectively of each X-ray source that
are brighter than the corresponding $K_s$-band completeness limit.

We represent the redshift density function for each cluster using both
a standard histogram ($\delta z = 0.1$) and a variable kernel density
estimation (VKDE) approach. The VKDE approach represents the
contribution of each galaxy to the redshift density function as a
Gaussian of width $\sigma = 0.03 \times (1+z)$ and unit area.
A first estimate of the photometric redshift of each cluster is
determined by identifying visually the peak associated with the X-ray
source in Figure \ref{fig:3panel}. We then compute the cluster
photometric redshift as the mean of the VKDE-weighted redshift
distribution, i.e.
\begin{equation}
{
z_{cluster}=\frac{\int z~K(z)~{\rm d} z}{\int K(z)~{\rm d} z}
}
\end{equation}
over the local interval where the VKDE distribution $K(z)$ exceeds
$0.5 \times K(z_{peak})$. The resulting $z_{cluster}$ for all
candidate clusters is displayed in Table \ref{tab_clus}. For the five
spectroscopically confirmed clusters with NIR data coverage we can
compute the effective photometric redshift error of this method as
\begin{equation}
{
\sigma_z = \left [ \frac{1}{N} \sum \left ( \frac{(z_{spec}-z_{phot})}{(1+z_{spec})} \right )^2 \right ]^{1/2}
}
\end{equation}
which yields $\sigma_z=0.025$. This error will naturally increase as
one extends this approach to the candidate clusters which typically
represent less clear redshift peaks composed of smaller numbers of
galaxies. Figure \ref{fig:3panel} also displays photometrically
selected cluster galaxies in field of each cluster. Cluster galaxies
are selected as occupying the local interval $K(z) > 0.5 \times
K(z_{cluster})$.  Figure \ref{fig:3panel} further displays the $J-K_s$
CMD for photometrically selected members of each cluster.

\begin{figure*}
\centering
\psfig{figure=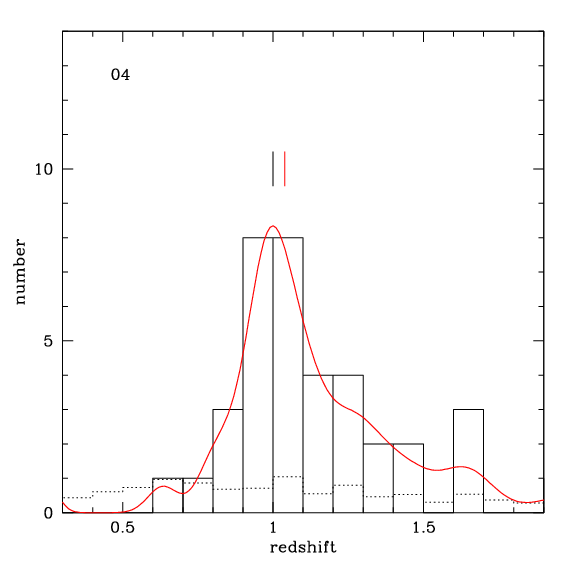,width=2.2in,angle=0.0}
\psfig{figure=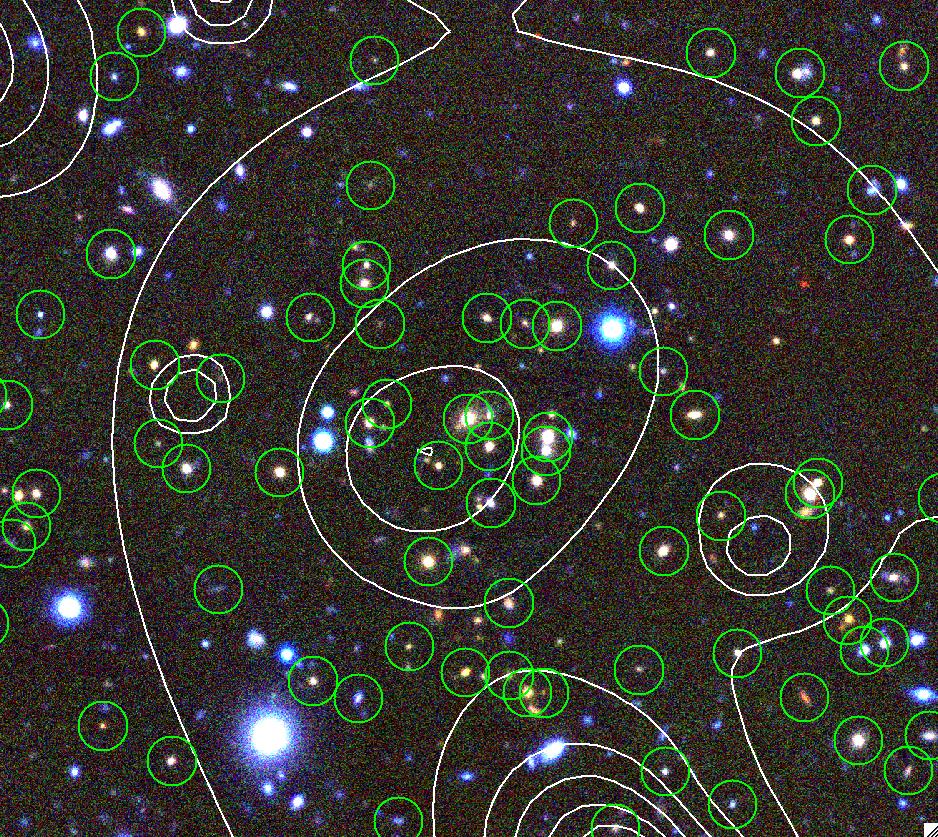,width=2.1in,angle=0.0}
\psfig{figure=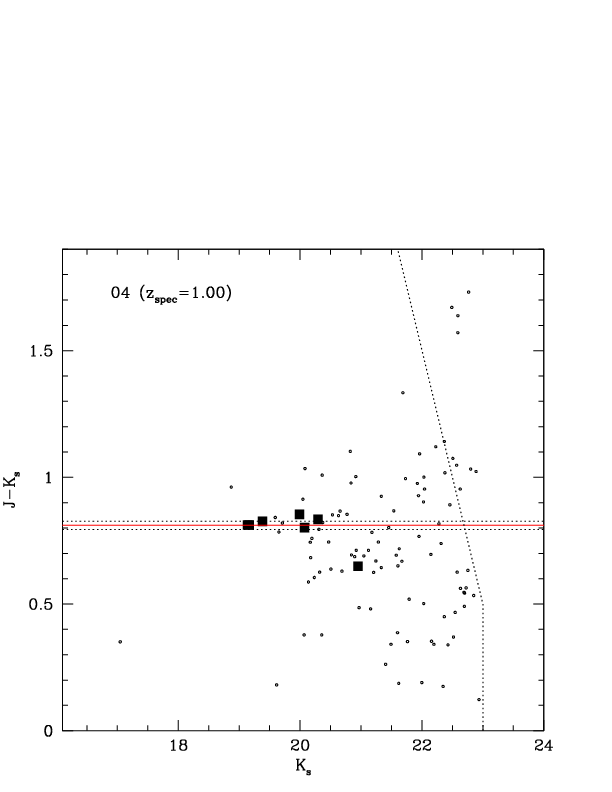,width=2.2in,angle=0.0}
\psfig{figure=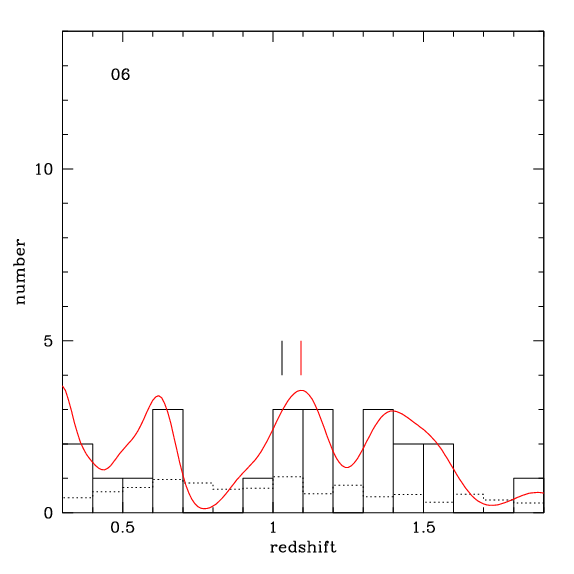,width=2.2in,angle=0.0}
\psfig{figure=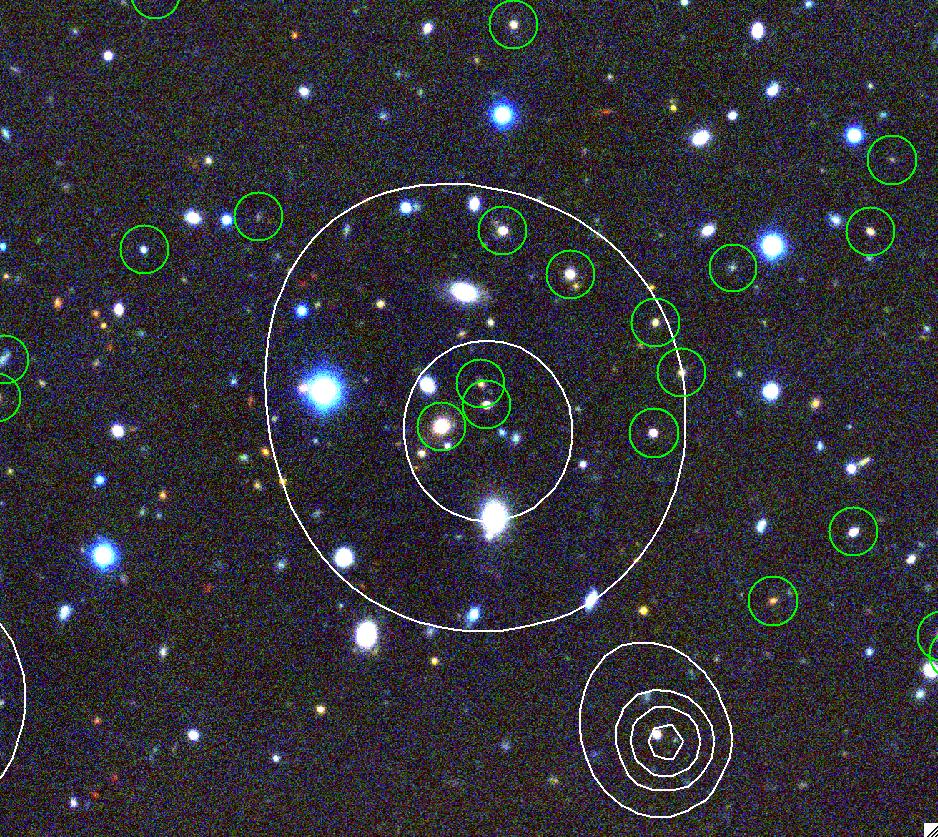,width=2.1in,angle=0.0}
\psfig{figure=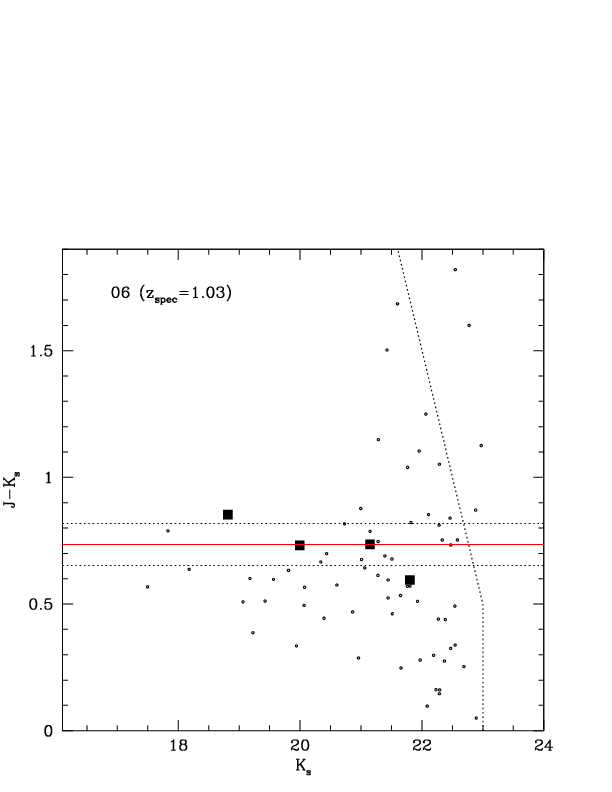,width=2.2in,angle=0.0}
\caption{Photometric redshifts and NIR photometry for each cluster
  with available $JK$ data (see text for additional information). Left
  panel: photometric redshift histogram for all sources located close
  to each X-ray cluster centroid. The red curve indicates the VKDE
  distribution and the dotted line is the appropriately scaled
  background. The vertical red tick mark indicates the computed
  photometric redshift for each cluster. For spectroscopically
  confirmed clusters the vertical black tick mark indicates the
  spectroscopic redshift. Middle panel: $iJK$ image of each
  cluster. Each image is 2\arcmin\ on a side and follows standard
  astronomical orientation. The white contours indicate the X-ray
  emission in each field and the green circles indicate
  photometrically selected cluster members. Right panel: colour
  magnitude diagram for sources within 1\arcmin\ of each cluster X-ray
  centroid. For spectroscopically confirmed clusters the black squares
  indicate spectroscopically selected cluster members. For candidate
  clusters the black squares indicate photometrically selected
  members. Points indicate all other sources. The horizontal red and
  dotted black lines indicate the computed red sequence location and
  associated uncertainty. The angled dotted line indicates the
  photometric completeness in each field.}
\label{fig:3panel}

\end{figure*}

\begin{figure*}
\centering
\psfig{figure=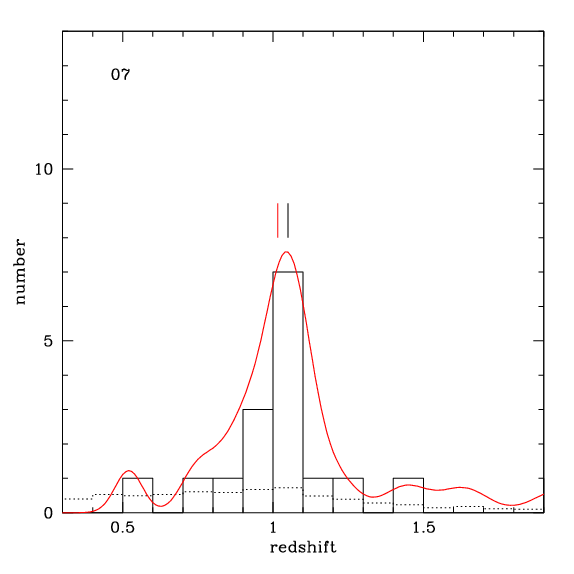,width=2.2in,angle=0.0}
\psfig{figure=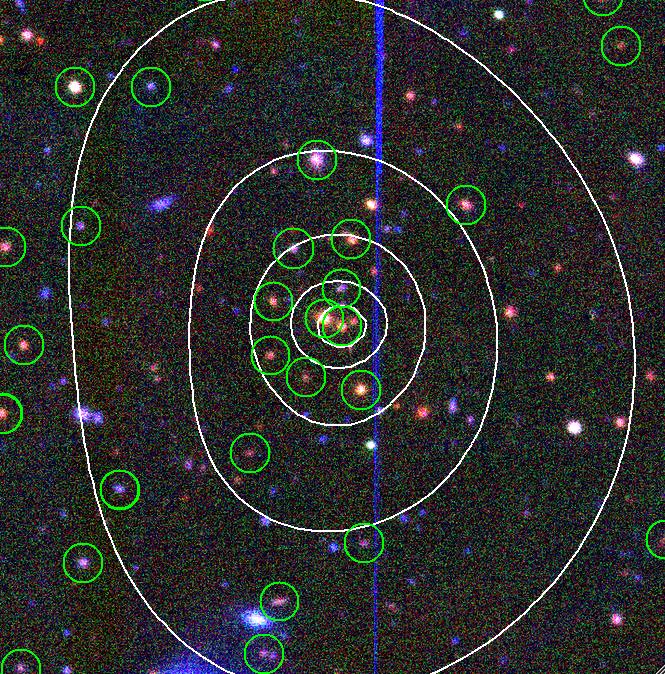,width=2.1in,angle=0.0}
\psfig{figure=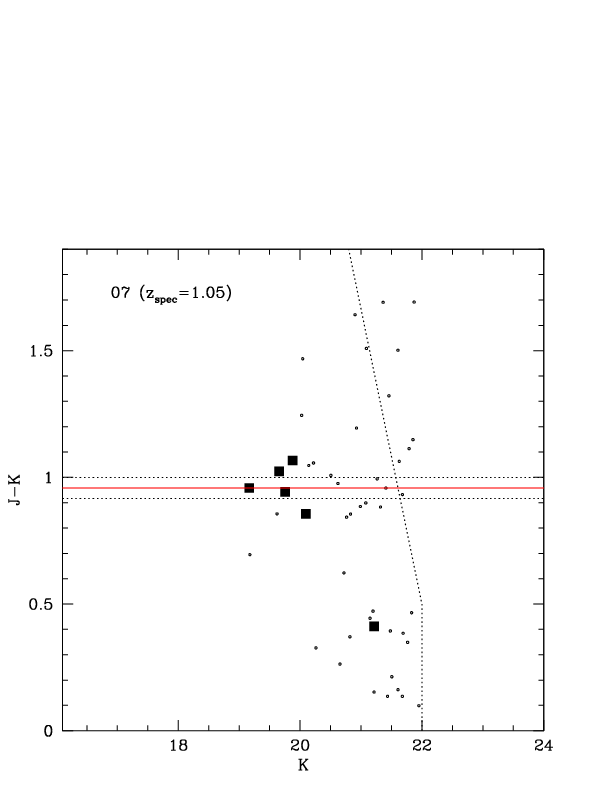,width=2.2in,angle=0.0}
\psfig{figure=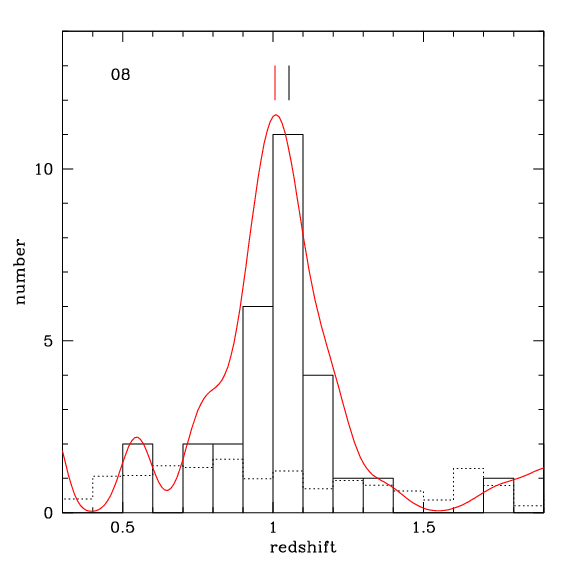,width=2.2in,angle=0.0}
\psfig{figure=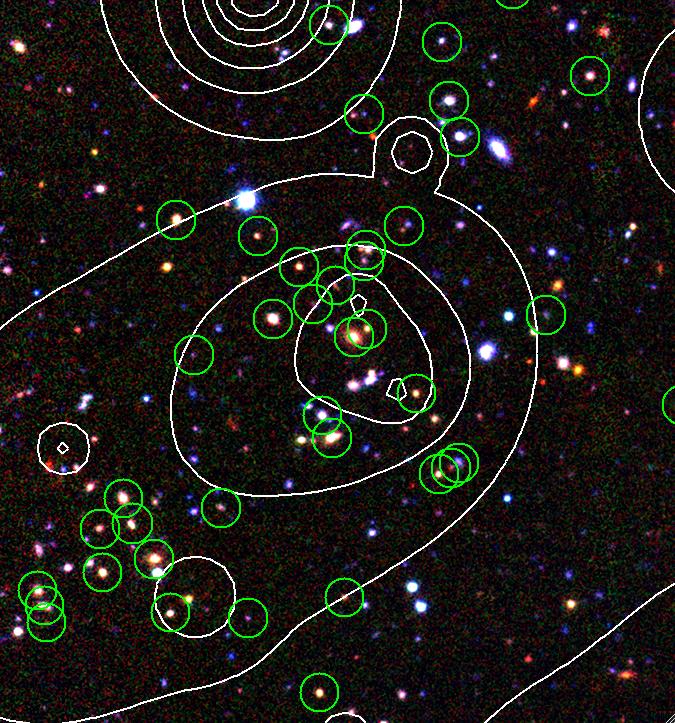,width=2.0in,angle=0.0}
\psfig{figure=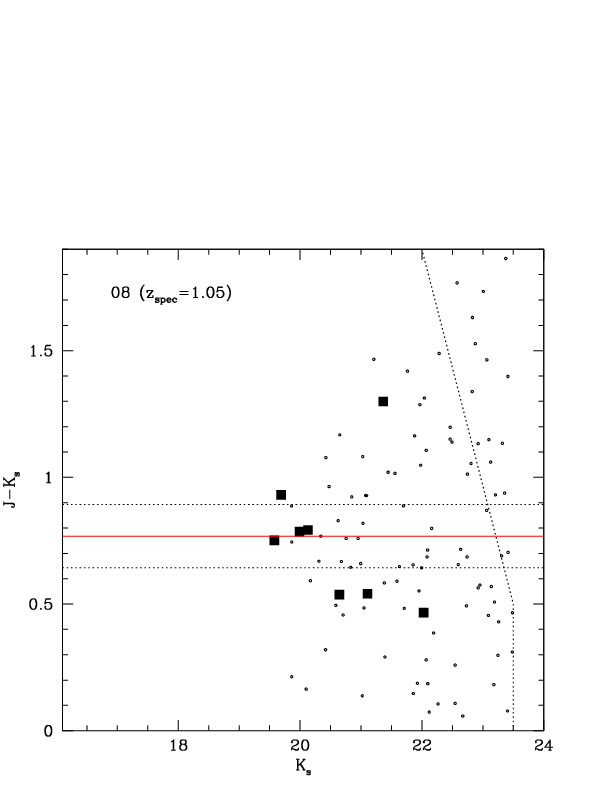,width=2.2in,angle=0.0}
\psfig{figure=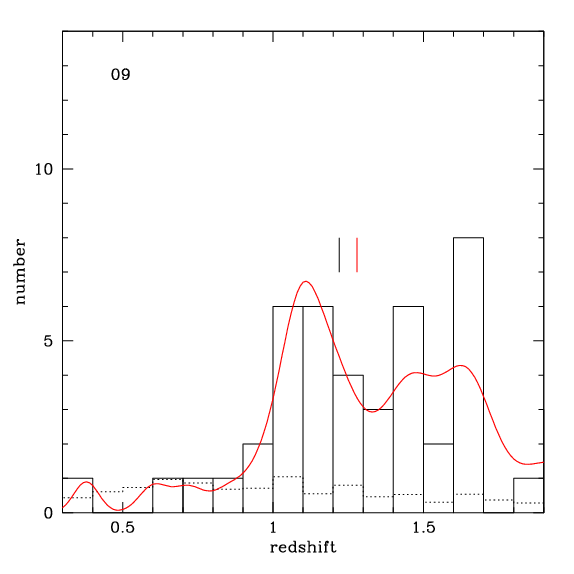,width=2.2in,angle=0.0}
\psfig{figure=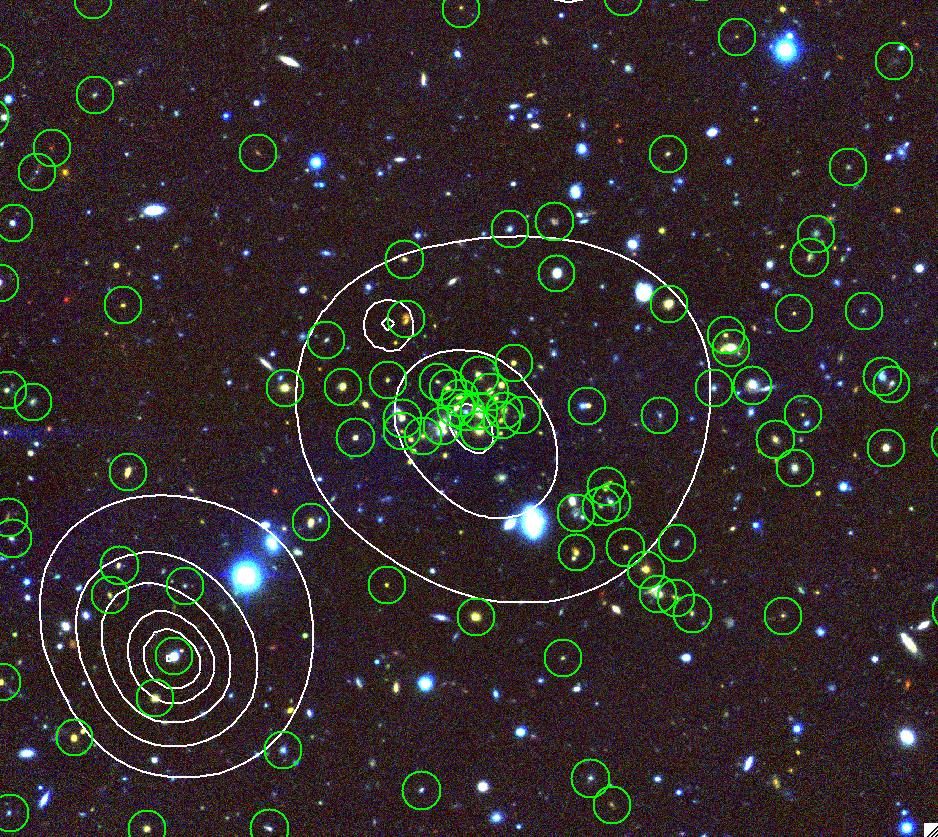,width=2.1in,angle=0.0}
\psfig{figure=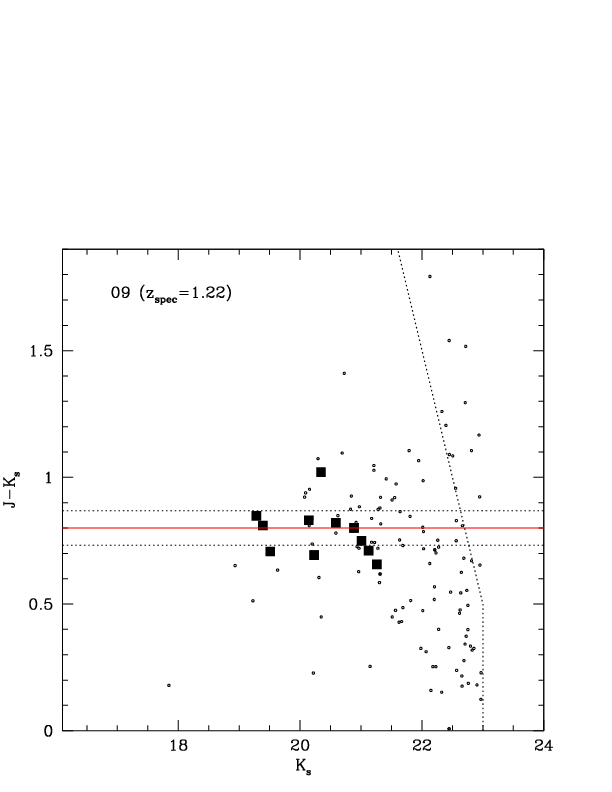,width=2.2in,angle=0.0}
\contcaption{}
\end{figure*}

\begin{figure*}
\centering
\psfig{figure=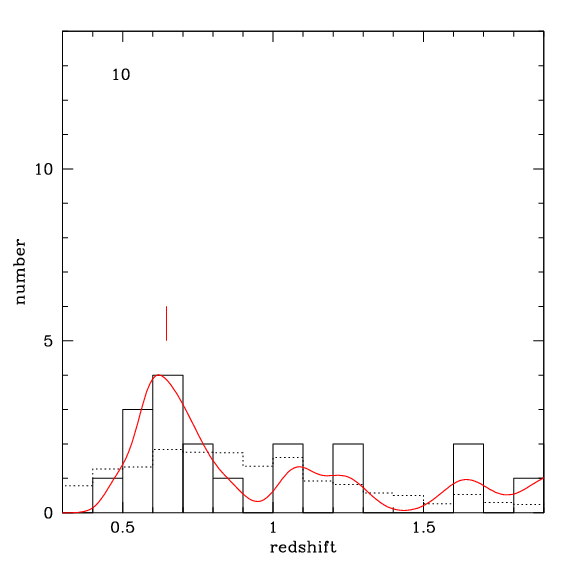,width=2.2in,angle=0.0}
\psfig{figure=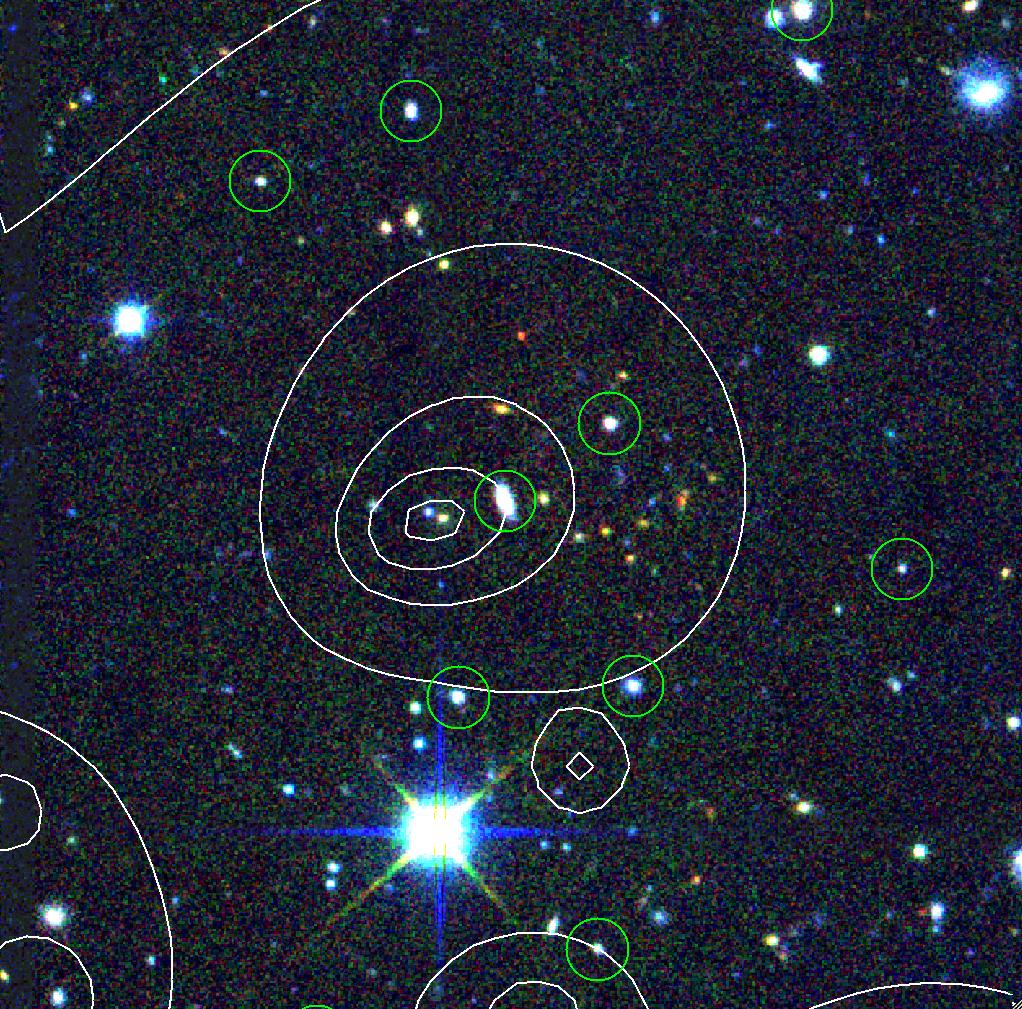,width=2.1in,angle=0.0}
\psfig{figure=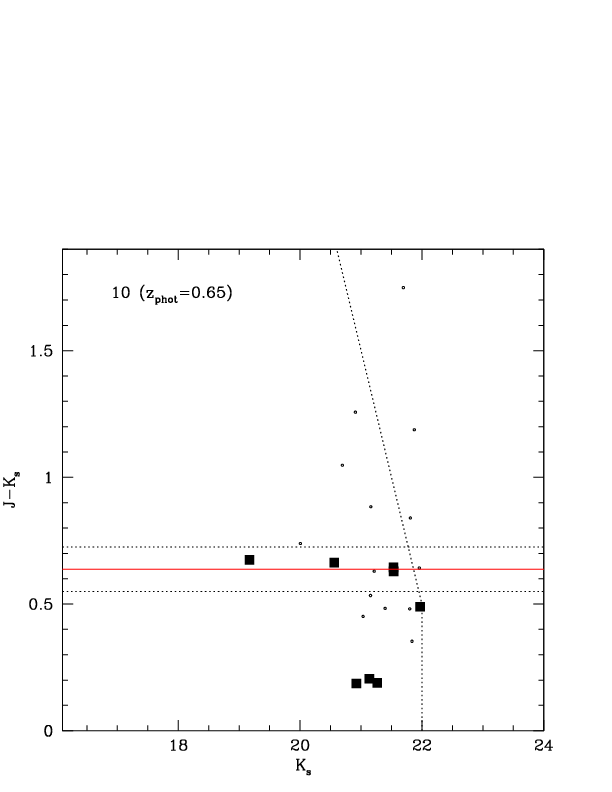,width=2.2in,angle=0.0}
\psfig{figure=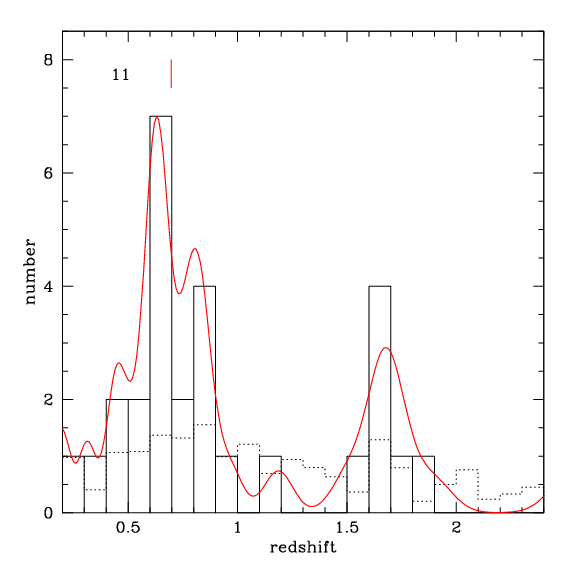,width=2.2in,angle=0.0}
\psfig{figure=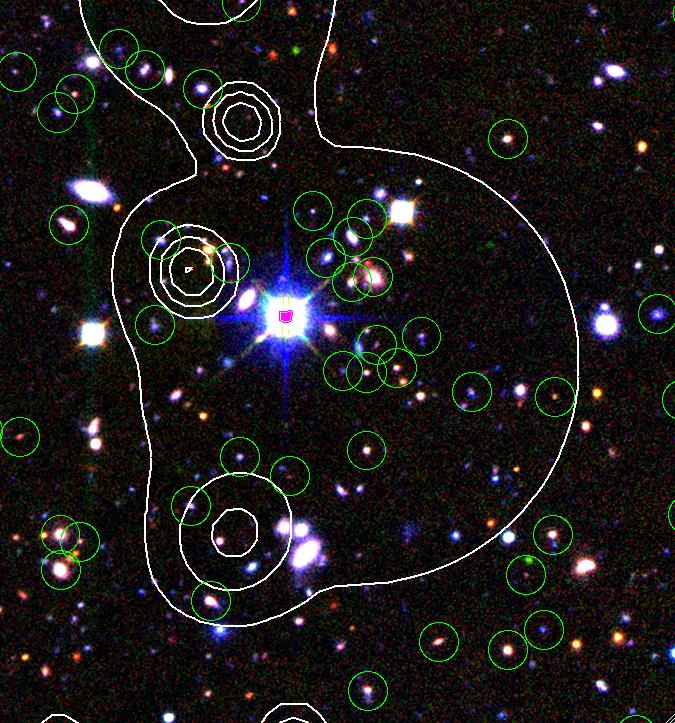,width=2.0in,angle=0.0}
\psfig{figure=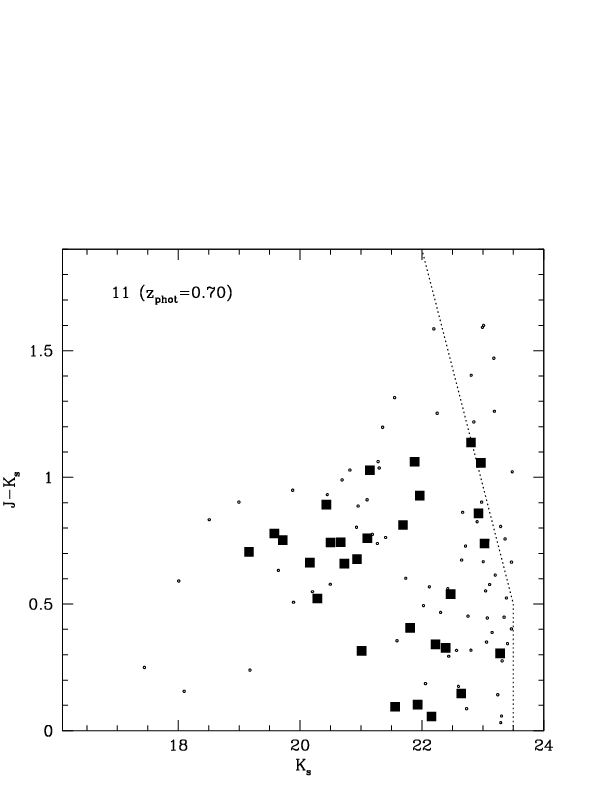,width=2.2in,angle=0.0}
\psfig{figure=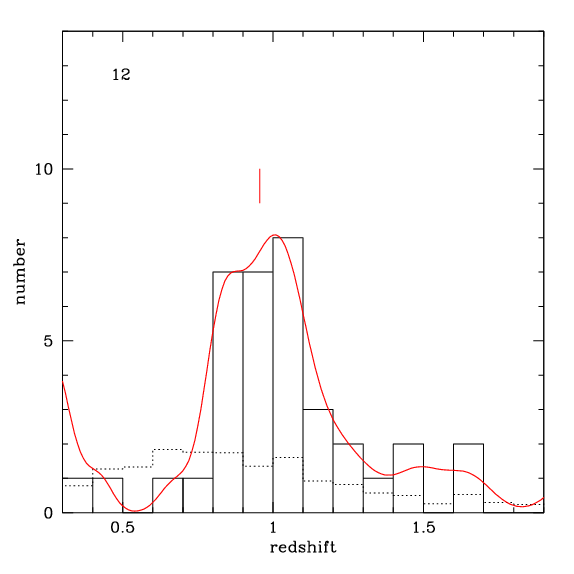,width=2.2in,angle=0.0}
\psfig{figure=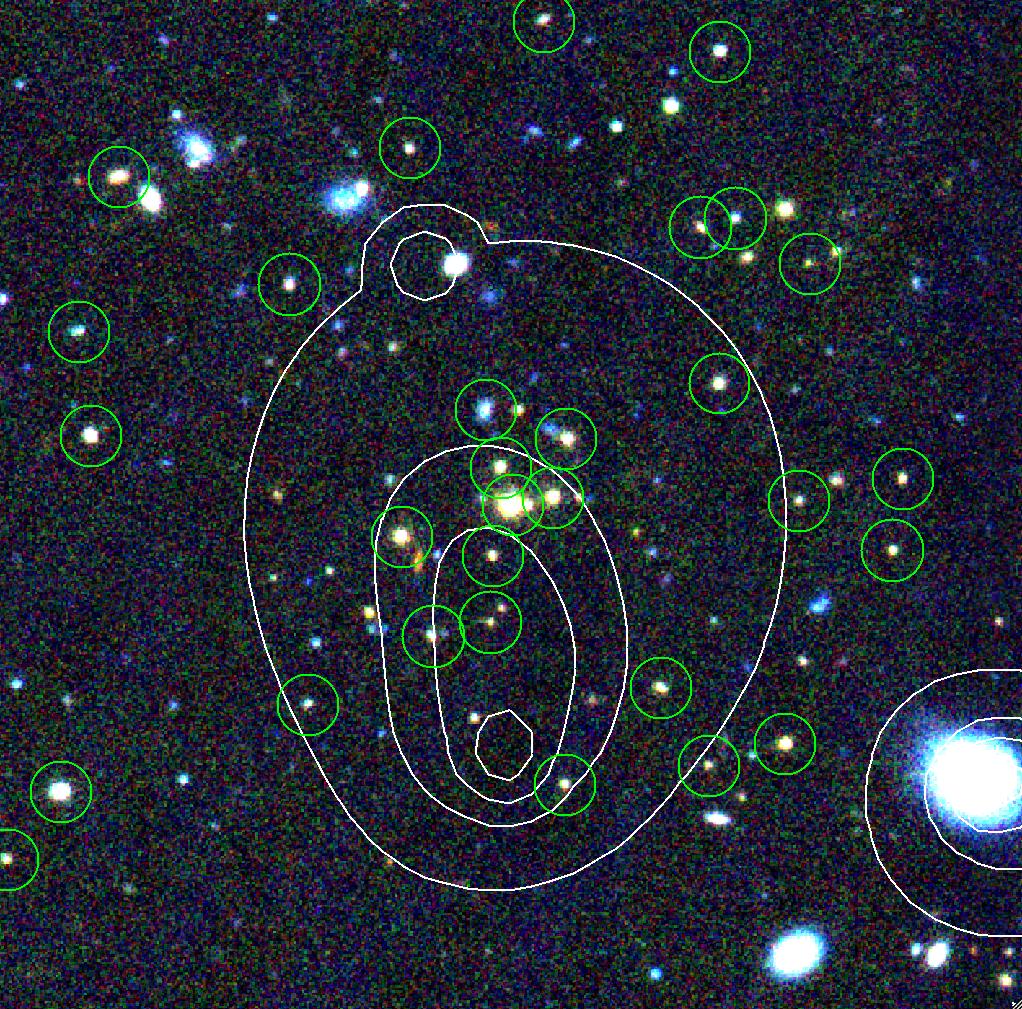,width=2.1in,angle=0.0}
\psfig{figure=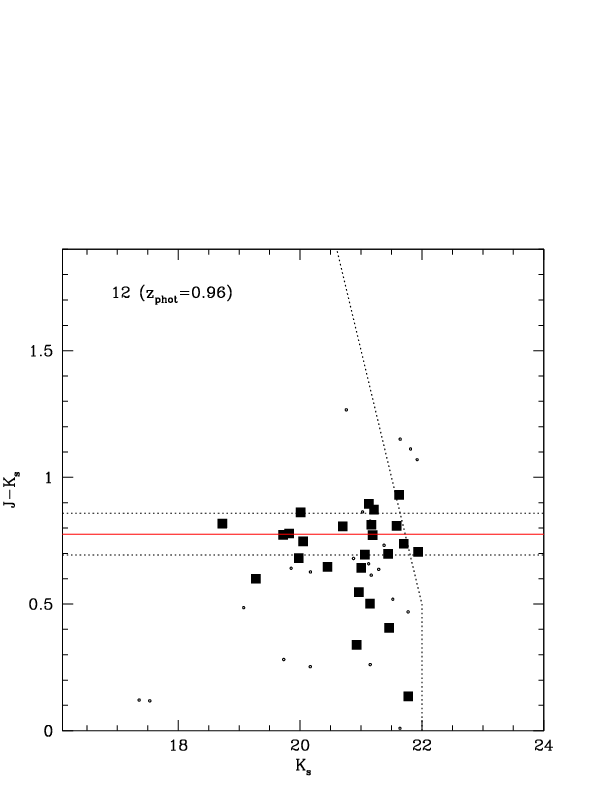,width=2.2in,angle=0.0}
\contcaption{}
\end{figure*}

\begin{figure*}
\centering
\psfig{figure=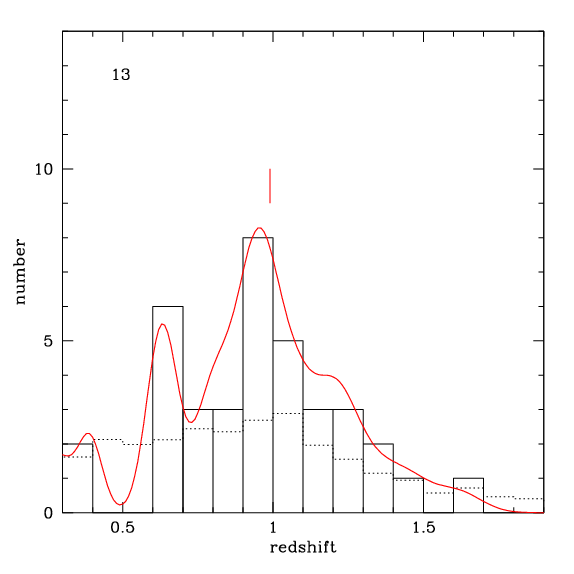,width=2.2in,angle=0.0}
\psfig{figure=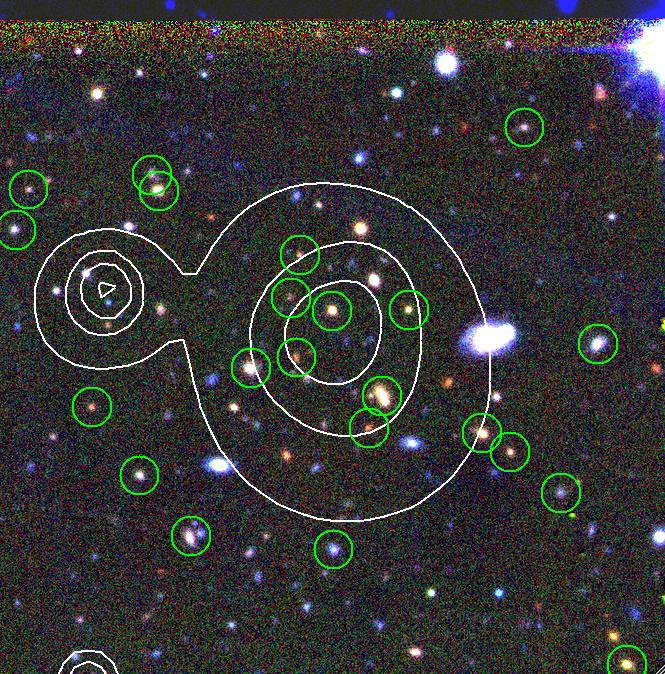,width=2.1in,angle=0.0}
\psfig{figure=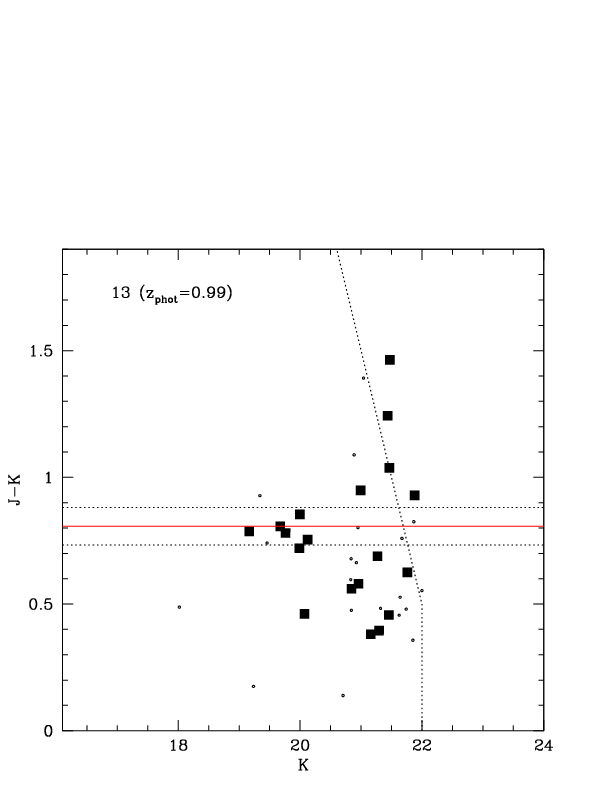,width=2.2in,angle=0.0}
\psfig{figure=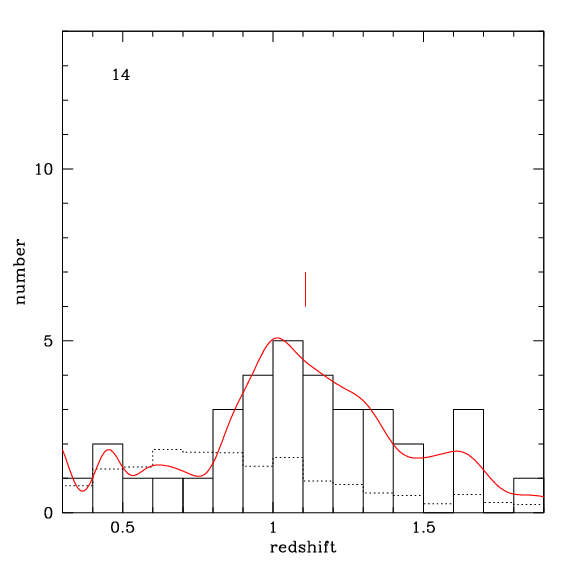,width=2.2in,angle=0.0}
\psfig{figure=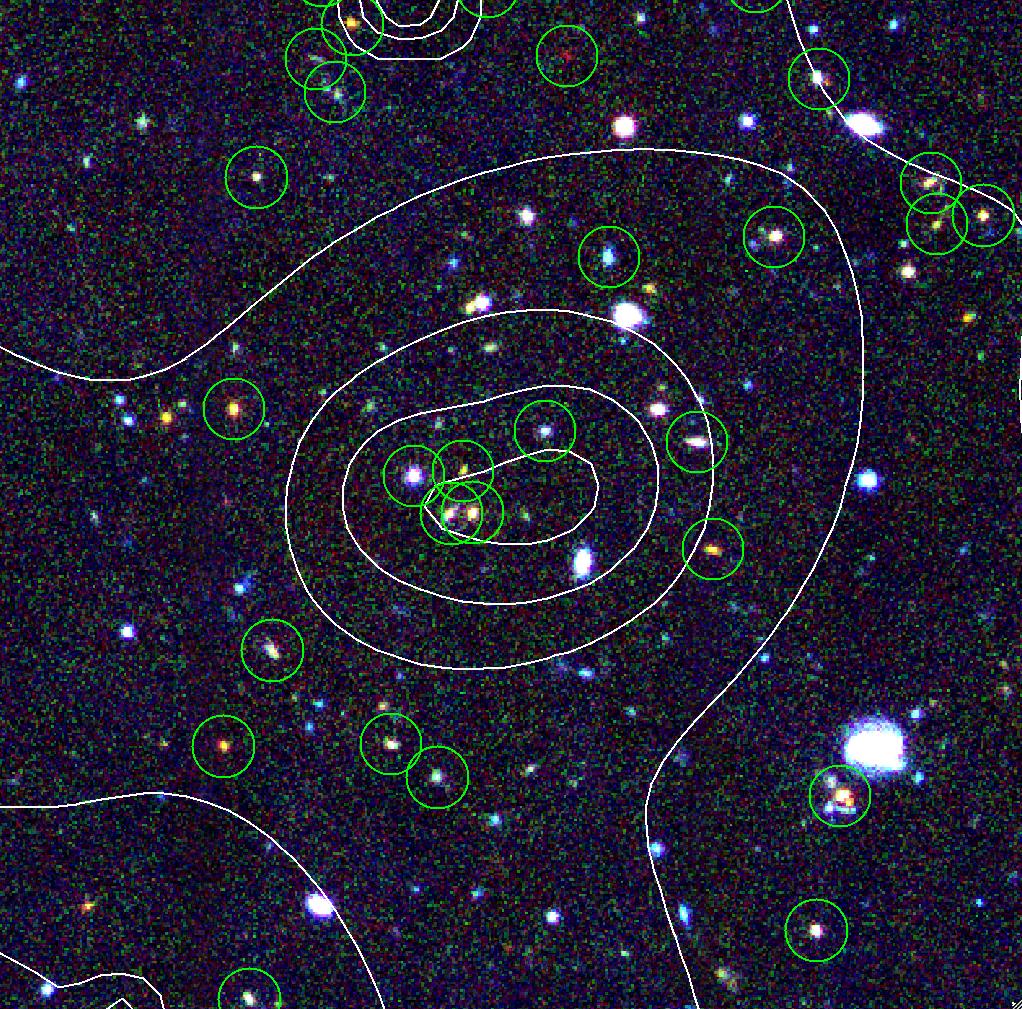,width=2.1in,angle=0.0}
\psfig{figure=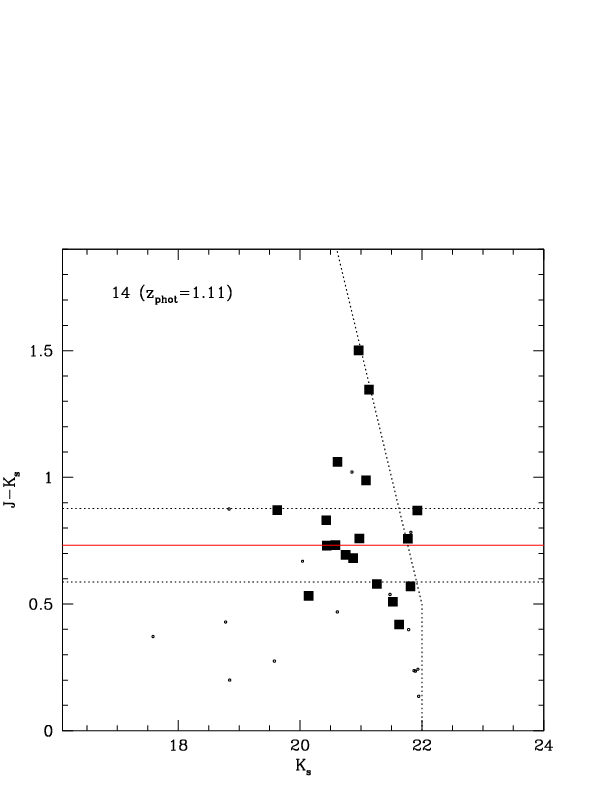,width=2.2in,angle=0.0}
\psfig{figure=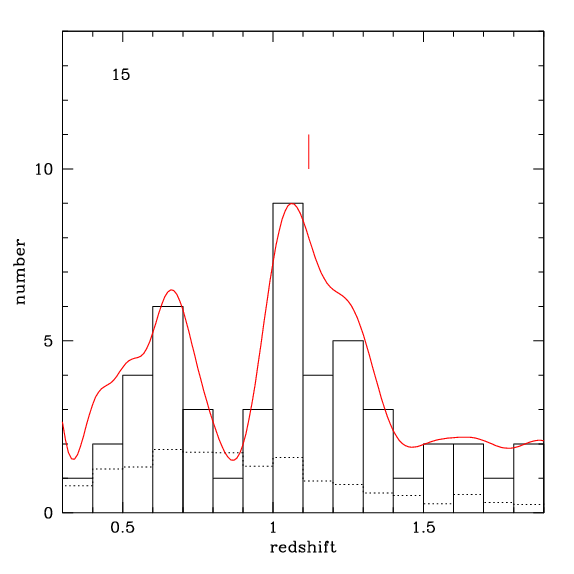,width=2.2in,angle=0.0}
\psfig{figure=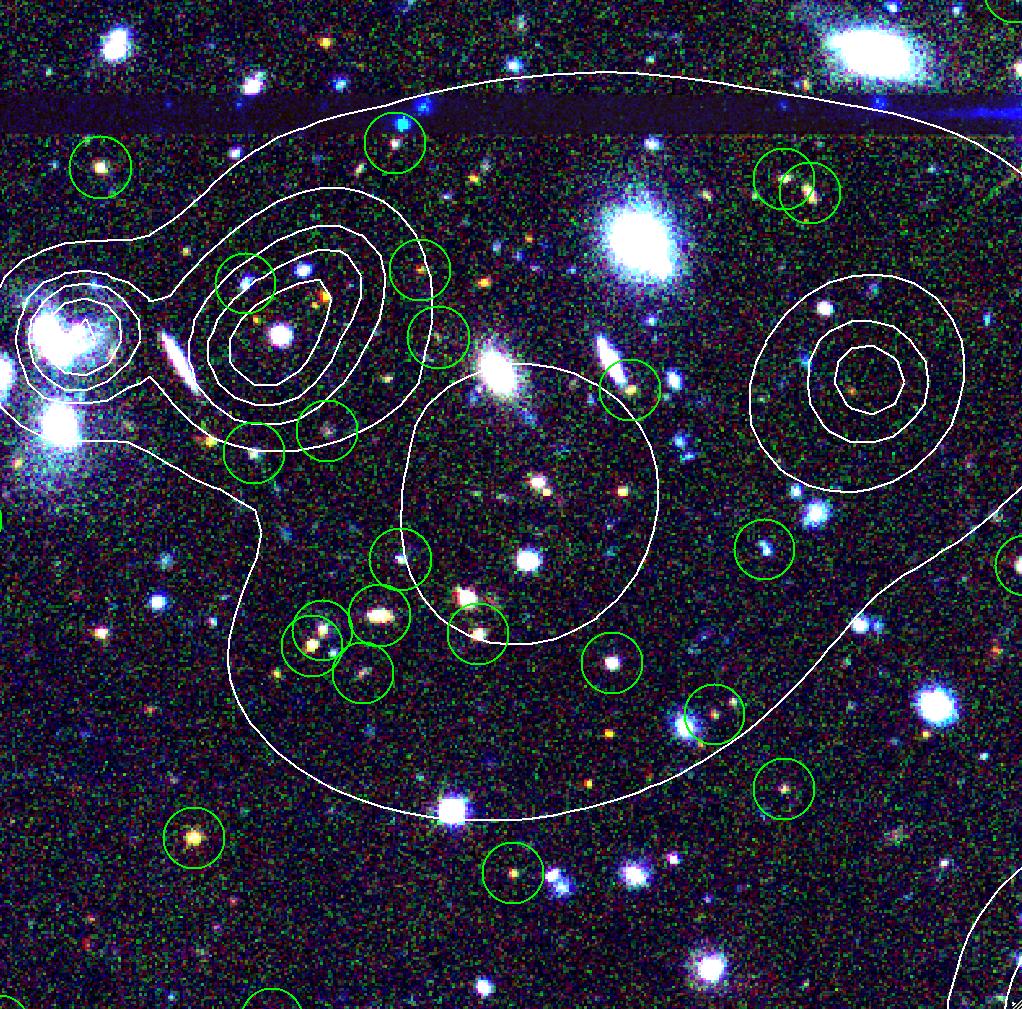,width=2.1in,angle=0.0}
\psfig{figure=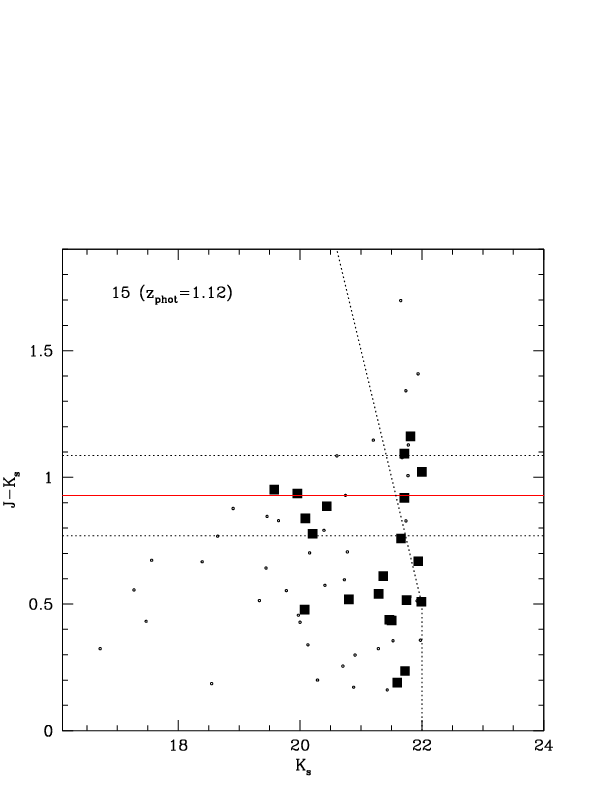,width=2.2in,angle=0.0}
\contcaption{}
\end{figure*}

\begin{figure*}
\centering
\psfig{figure=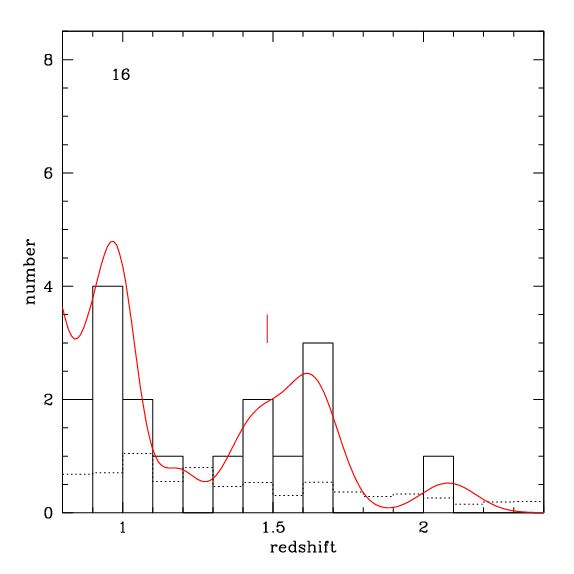,width=2.2in,angle=0.0}
\psfig{figure=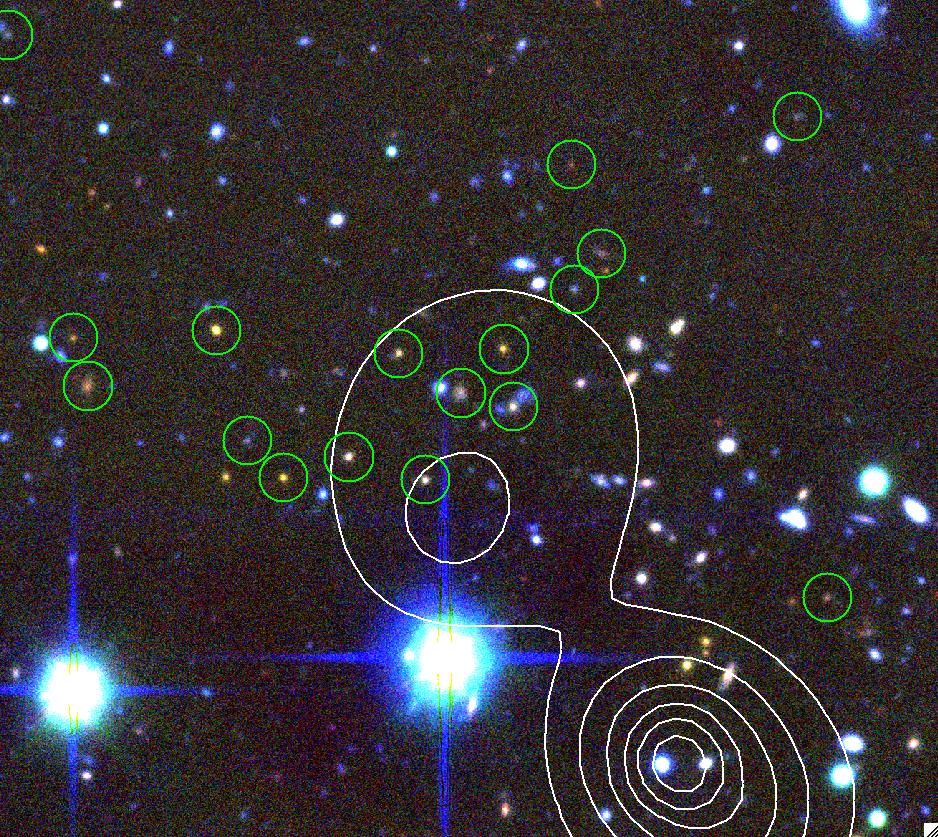,width=2.1in,angle=0.0}
\psfig{figure=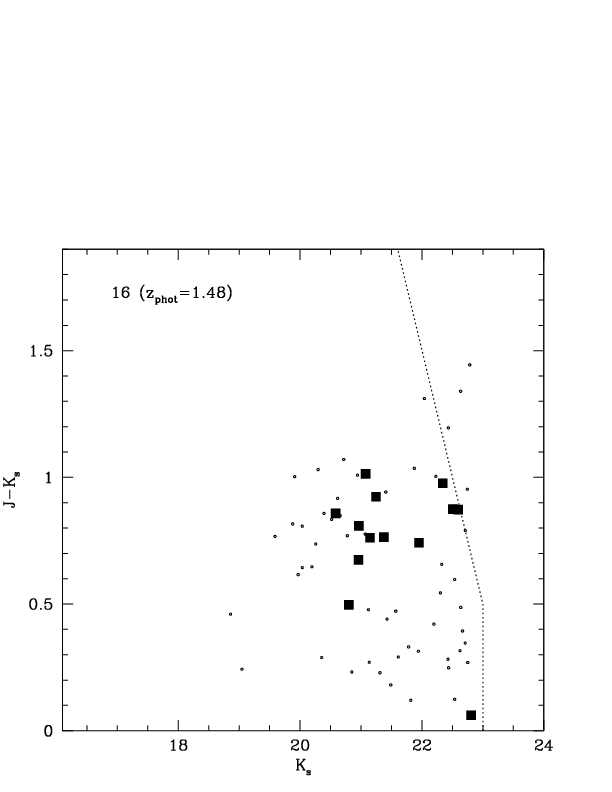,width=2.2in,angle=0.0}
\psfig{figure=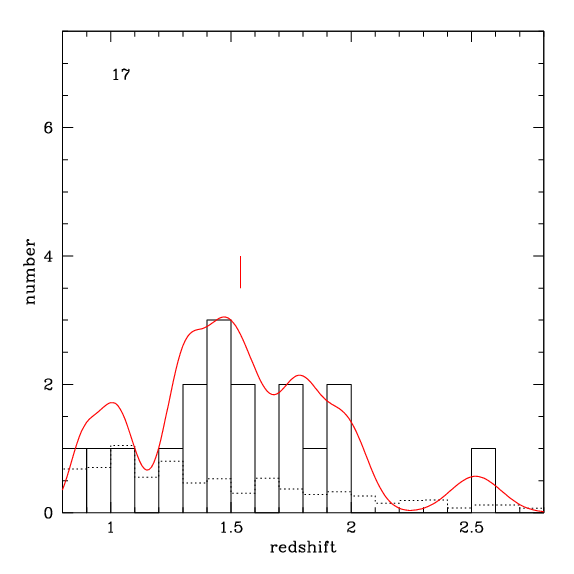,width=2.2in,angle=0.0}
\psfig{figure=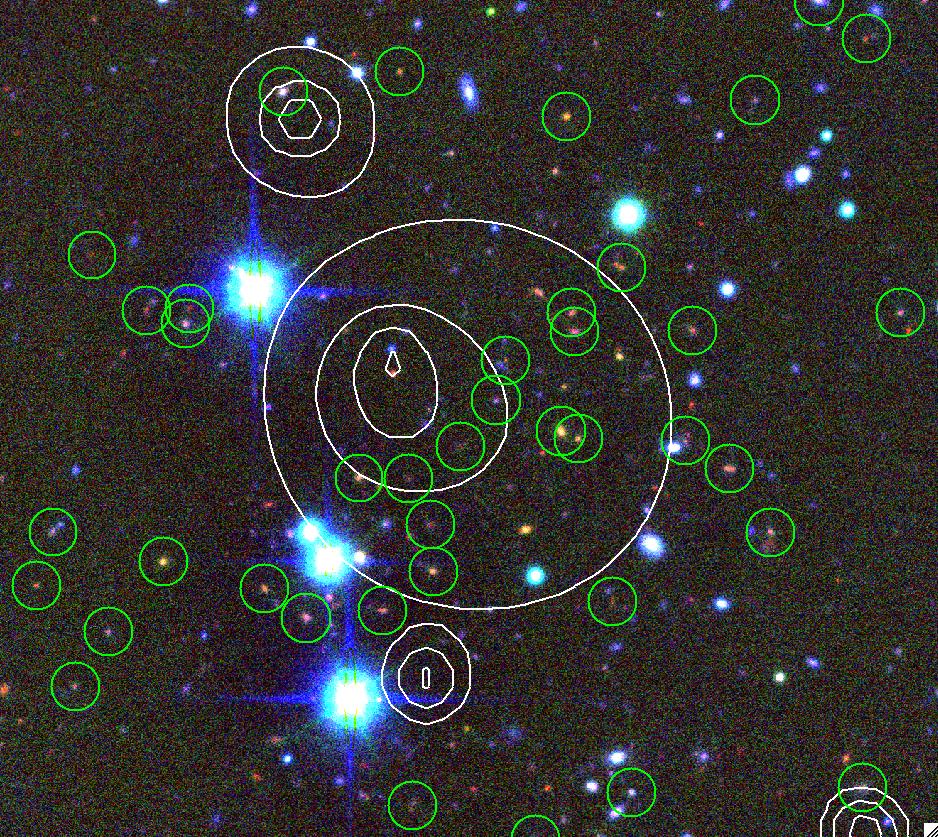,width=2.1in,angle=0.0}
\psfig{figure=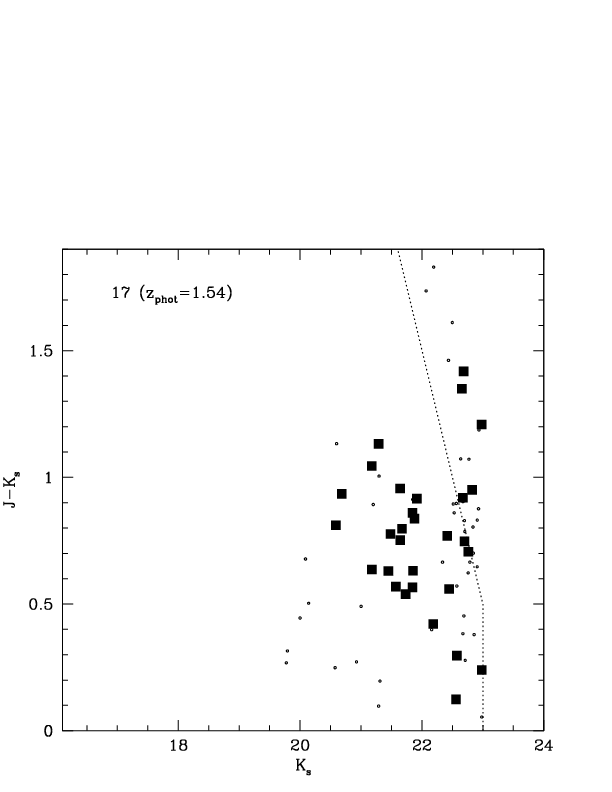,width=2.2in,angle=0.0}
\psfig{figure=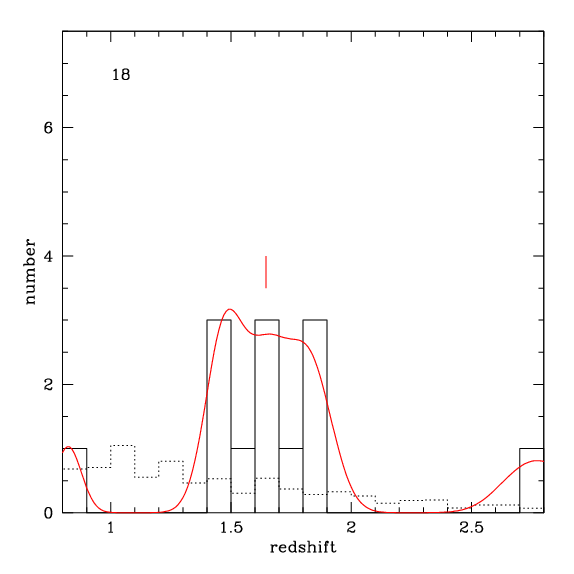,width=2.2in,angle=0.0}
\psfig{figure=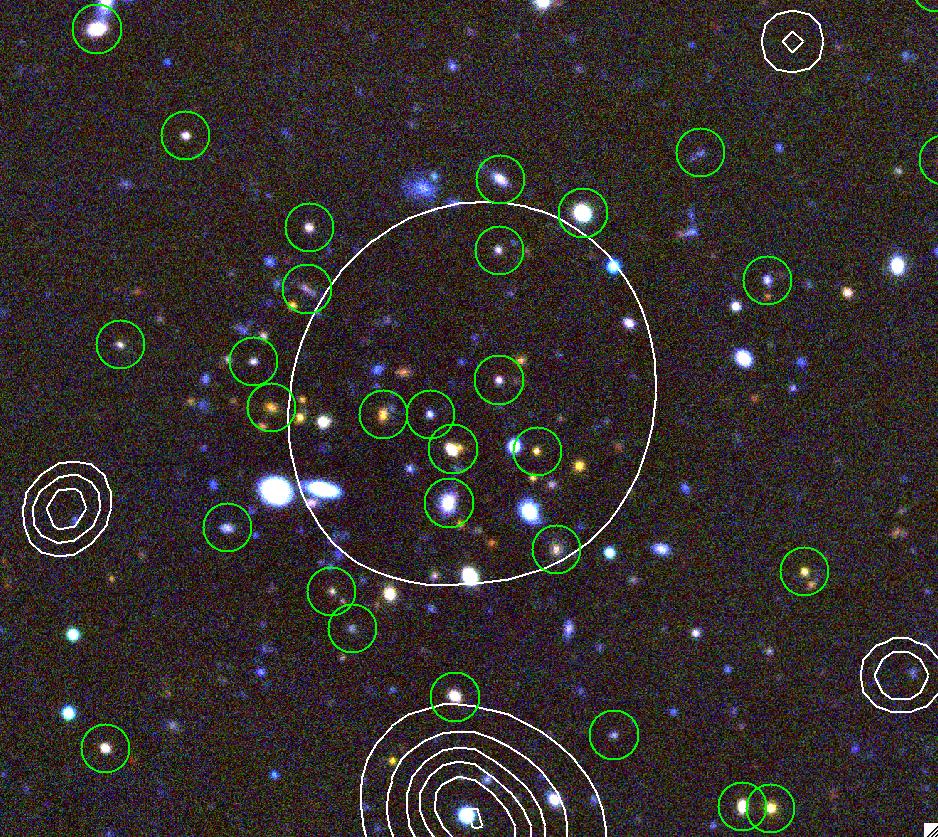,width=2.1in,angle=0.0}
\psfig{figure=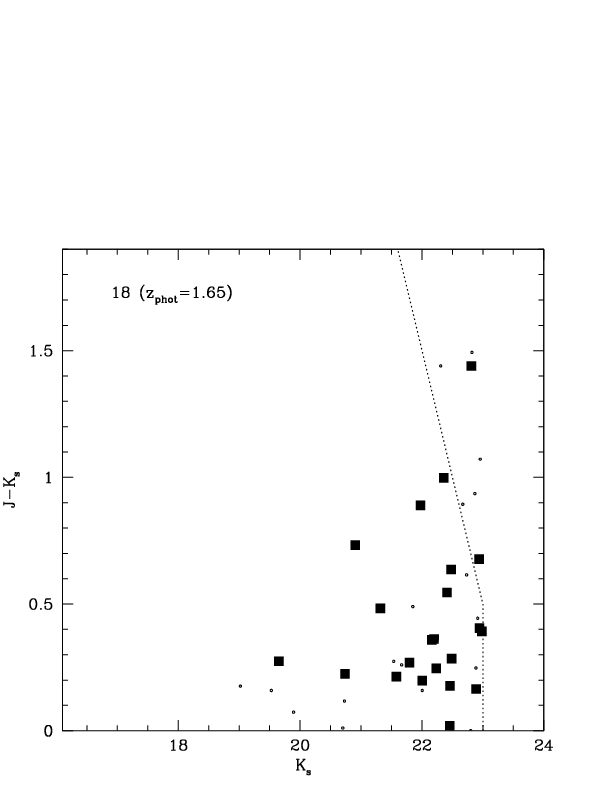,width=2.2in,angle=0.0}
\contcaption{}
\end{figure*}

\begin{figure*}
\centering
\psfig{figure=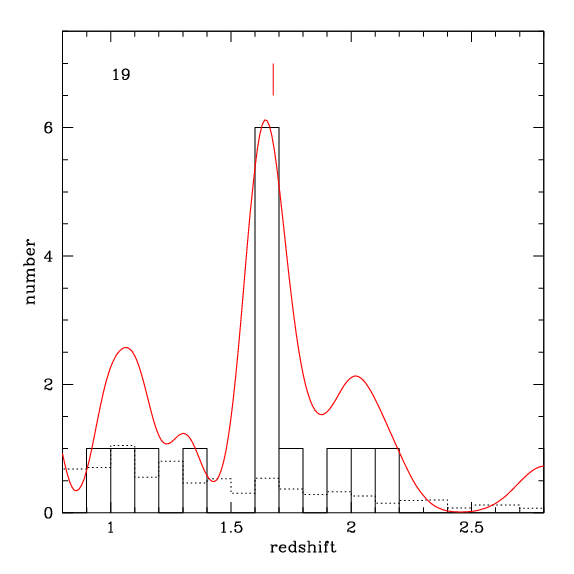,width=2.2in,angle=0.0}
\psfig{figure=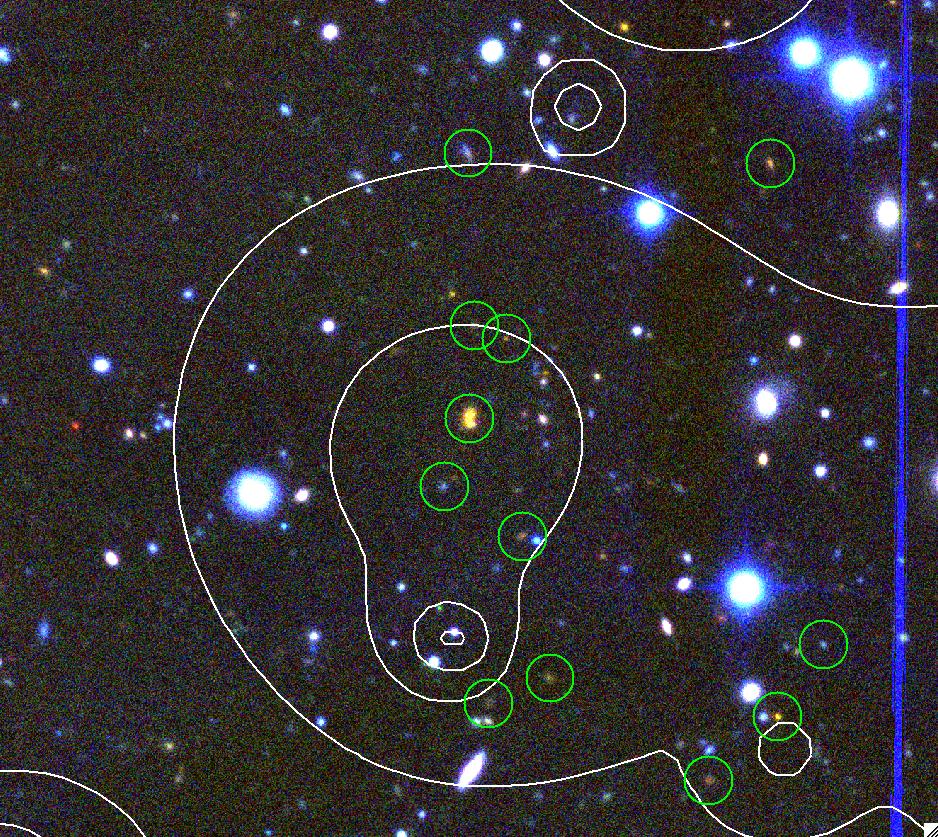,width=2.1in,angle=0.0}
\psfig{figure=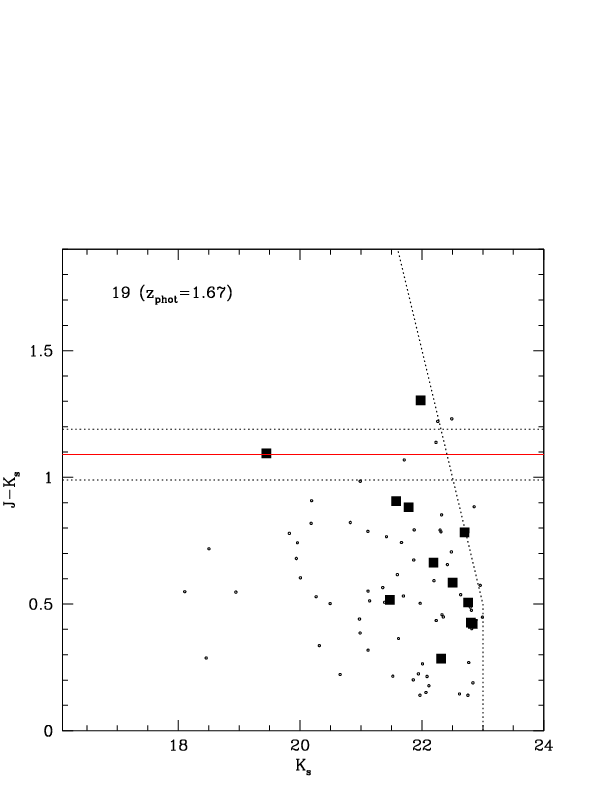,width=2.2in,angle=0.0}
\psfig{figure=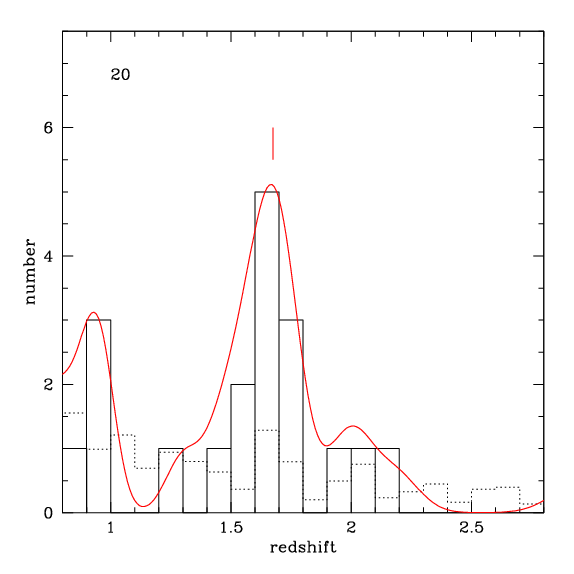,width=2.2in,angle=0.0}
\psfig{figure=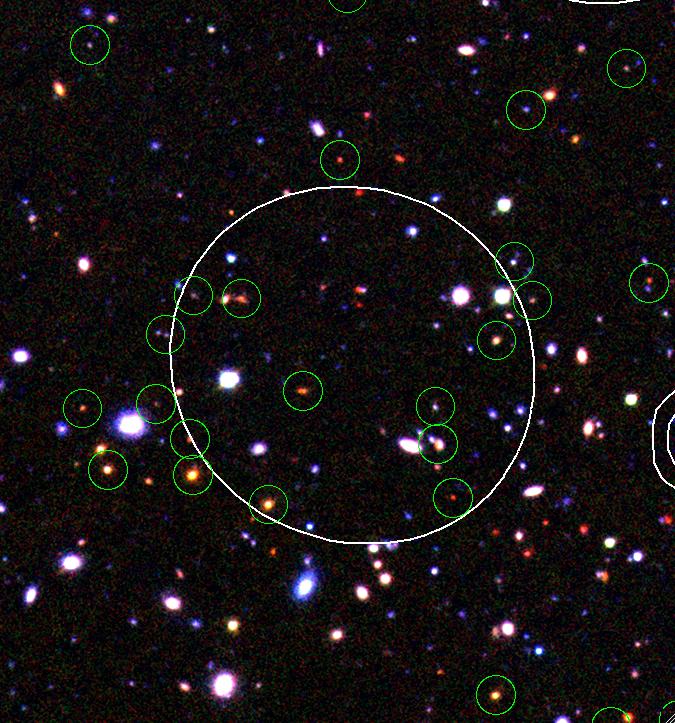,width=2.1in,angle=0.0}
\psfig{figure=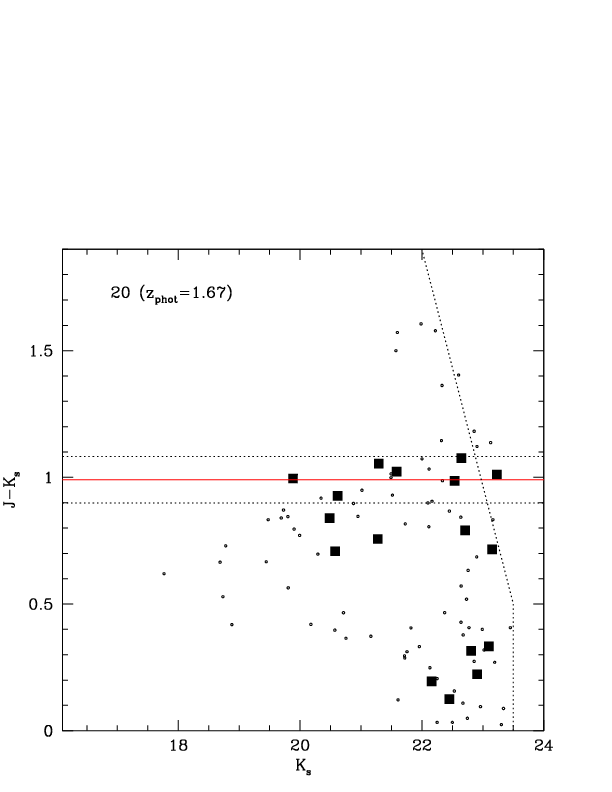,width=2.2in,angle=0.0}
\psfig{figure=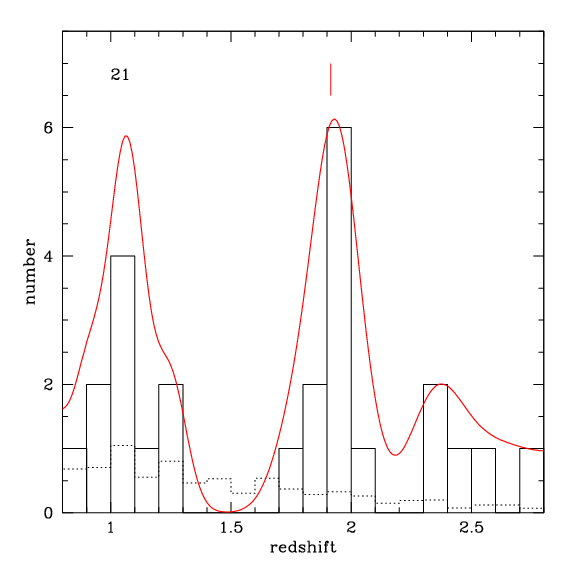,width=2.2in,angle=0.0}
\psfig{figure=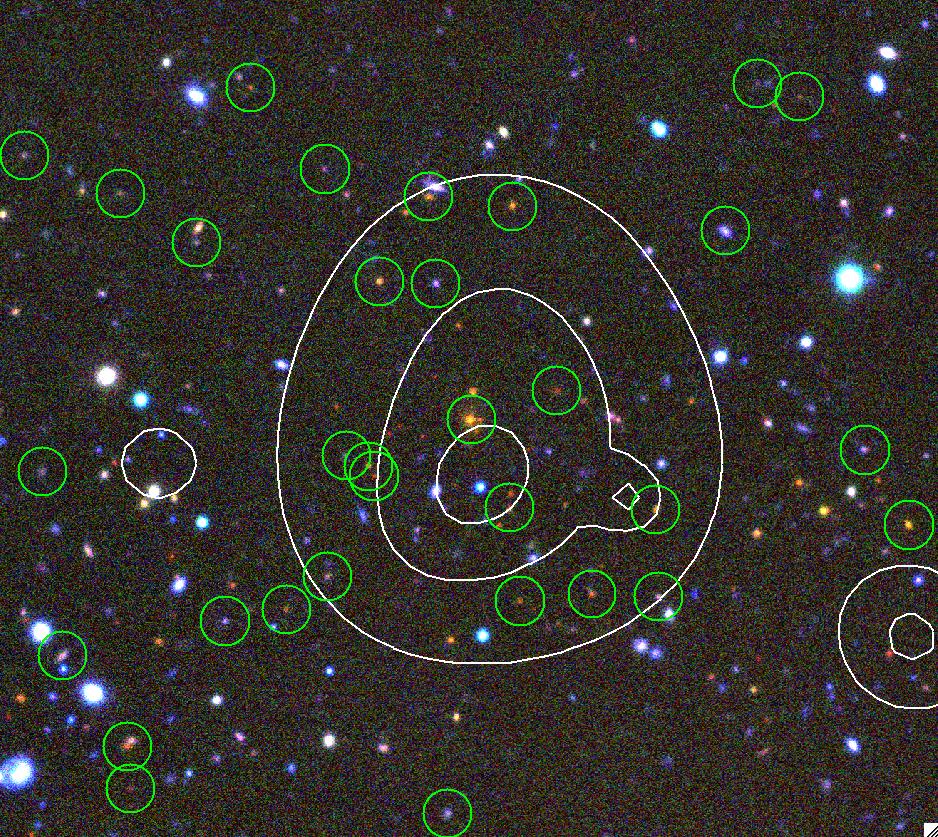,width=2.1in,angle=0.0}
\psfig{figure=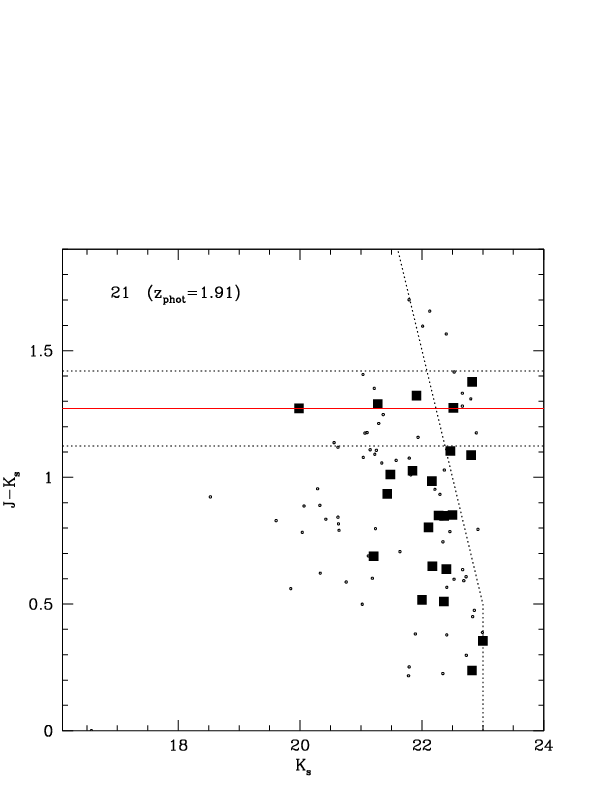,width=2.2in,angle=0.0}
\contcaption{}
\end{figure*}

\begin{figure*}
\centering
\psfig{figure=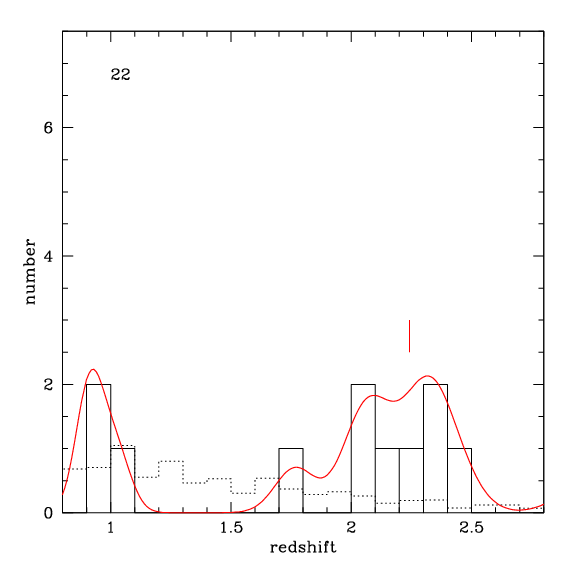,width=2.2in,angle=0.0}
\psfig{figure=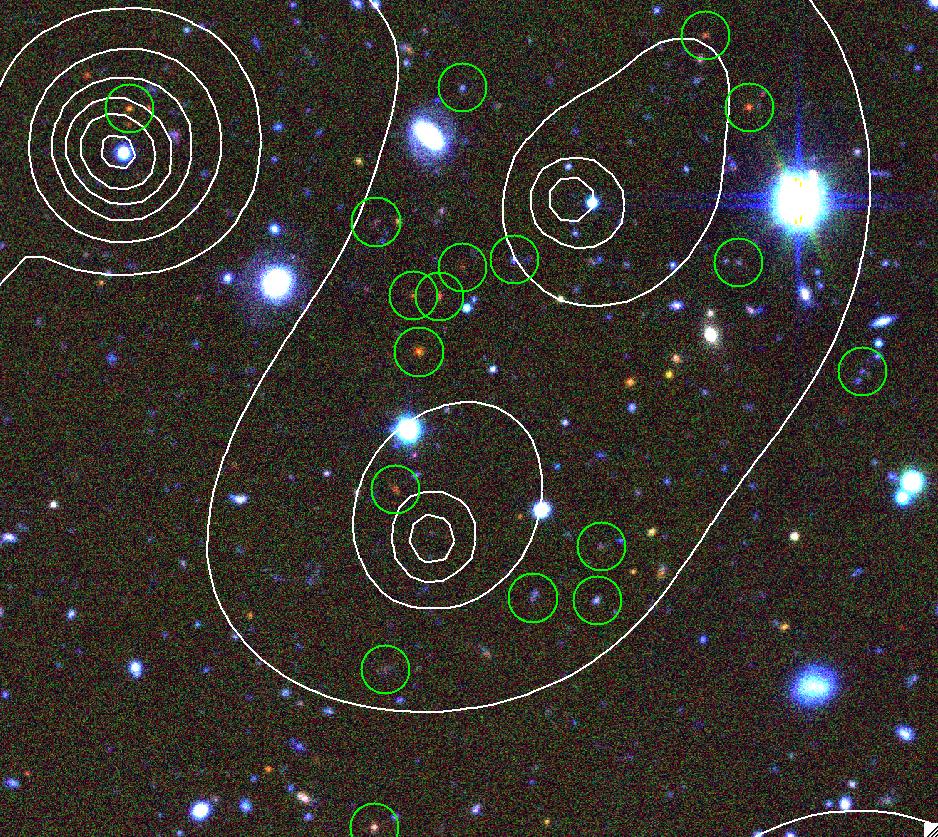,width=2.1in,angle=0.0}
\psfig{figure=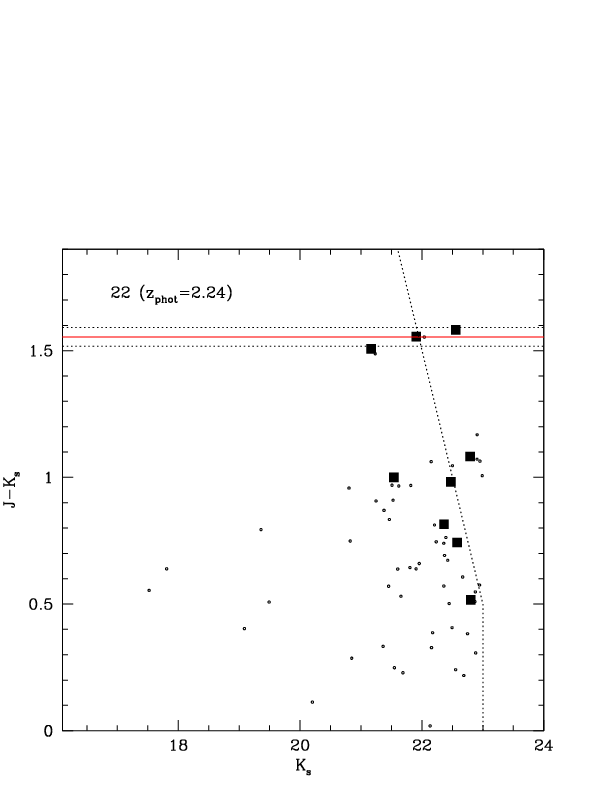,width=2.2in,angle=0.0}
\contcaption{}
\end{figure*}

\section{Discussion}
\label{discuss}

\subsection{Cluster red sequences: implications for galaxy assembly
  history}

In Section \ref{sec_rzI1} we investigated the limitations of the
$r-3.6\micron$ and $3.6\micron-4.5\micron$ data in the determination
of the location of cluster red sequences.  We return to this issue
with the NIR data and the results of the photometric redshift analysis
in hand and investigate two approaches to compute the colour sequence
of cluster member galaxies selected by a) photometric redshift and 2)
statistical background subtraction on the NIR colour magnitude plane.

\subsubsection{Computing the red sequence colour employing photometric redshifts}

\begin{figure}
\centering
\psfig{figure=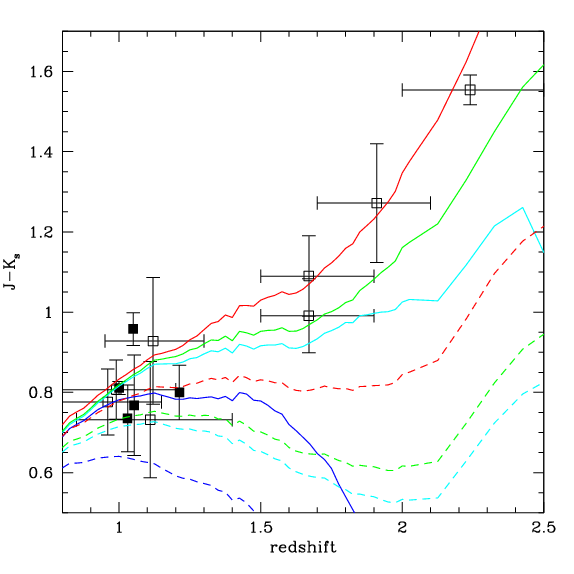,width=3.5in,angle=0.0}
\caption{Cluster red sequence colour computed for spectroscopically
  confirmed (solid squares) and candidate (open squares) clusters.
  The colour-redshift data are compared to model predictions generated
  using the BC03 spectral sysnthesis code \citep{bc03}. Each model
  features either a 1 Gyr burst (solid line) or $\tau=1$ Gyr (dashed
  line) solar metallicity stellar population with a standard Salpeter
  IMF. Models are displayed for formation redshifts of 10 (red), 5
  (green), 4 (cyan) and 3 (blue) respectively. }
\label{fig_jmkred}

\end{figure}

Figure \ref{fig:3panel} displays the CMDs of cluster galaxies in
confirmed and candidate clusters selected by photometric redshift (or
spectroscopic redshift if available) and proximity to the X-ray
source.  In order to identify the location of any red sequence in each
cluster we apply the method outlined by \citet{fass11} whereby one
identifies the third reddest galaxy in each CMD (3RG) and selects
galaxies displaying $J-K_s$ colours within $\rm 3RG \pm 0.3$. We then
compute the cluster ``red sequence'' colour as the median of this set
of galaxies ($C_{med}$) and the spread as either the colour of the
galaxies enclosing 68\% of the distribution (i.e. $\sigma_c=0.5 \times
[C_{84}-C_{16}]$; for $N>3$) or the full colour range (for $N \le
3$). We apply this method to all confirmed and candidate clusters with
$J-K_s$ imaging and display the resulting $C_{med}$ and $\sigma_c$
values in Figure \ref{fig_jmkred}. For cluster 19 we set $C_{med}$
equal to the colour of the candidate BCG and $\sigma_c=0.1$.  In many
cases this approach identifies a viable, yet often poorly populated,
red sequence. However, we note that this approach does not identify a
clear red sequence for candidate clusters 16, 17 and 18. We address
this result in more detail in the Section \ref{backsub}.

For those clusters displaying a clear red sequence consistent
with $C_{med}$ and $\sigma_c$ we compare the red sequence colour
versus redshift data to a set of representative colour-redshift loci
generated using the BC03 spectral synthesis code \citep{bc03}. The
plot indicates that the location of the putative red sequences of the
confirmed and candidate clusters follow a clear
locus in colour-redshift space that can be described by the simple
passive evolution of an old stellar population. We do not
attempt to fit the stellar population parameters best describing the
data beyond noting that solar metallicity models arising from a
1 Gyr burst of star formation at $z_f>5$ appear to be favoured over
either extended ($\tau=1$ Gyr), sub-solar ($Z=0.4Z_\odot$) or younger
($z_f<5$) bursts of star formation. It is worth noting at this point
that the strength of this conclusion rests upon the red sequence
colours of the candidate clusters at $z>1.5$. In one sense this
demonstrates the important leverage that distant clusters place upon
our knowledge of galaxy evolution in dense environments. In another
sense however, this result can only be viewed as tentative pending
spectroscopic confirmation of these clusters.

\subsubsection{Computing the red sequence location employing background subtraction on the CMD plane}
\label{backsub}

An alternative approach to identifying the red sequence of candidate
cluster galaxies is to investigate their distribution directly on the
colour magnitude plane. The main issue is to isolate the potentially
weak cluster signature from the ``background'' of non-cluster galaxies
along the line of sight.  To achieve this we bin the colour magnitude
distribution of galaxies located within 1\arcmin\ of each cluster
X-ray centroid. Galaxies are binned on the $J-K_s$ versus $K_s$ plane with
bin dimensions of 0.2 and 0.5 mag. respectively.  The CMD of
non-cluster galaxies (termed the model background here) is formed by
selecting all sources at $>1\arcmin$ from each cluster or cluster
candidate and scaling the resulting distribution by the relative
cluster and background sky areas.

Although the resulting background subtracted CMD for each cluster
displays an identifiable overdensity of faint, red galaxies, the
signal is accompanied by variations in the true background associated
with each cluster (both Poisson and cosmic in nature). Upon
subtraction of the scaled model background these variations persist as
residual positive and negative signatures. However, their statistical
distribution should average to zero over all cluster fields and, in an
attempt to reduce their impact, we stack the binned subtracted CMDs to
investigate the average cluster CMD.  Prior to stacking we shift each
cluster distribution on the colour magnitude plane to account for the
$k$-correction and distance dimming of an evolving stellar population
at each cluster redshift.  We correct each cluster to a common
redshift applying an apparent colour and magnitude shift based upon
the evolution of a 1 Gyr solar metallicity burst of star formation
occuring at $z_f=10$ and described by a Salpeter IMF. Using this same
model we have confirmed that the colour terms between the different
NIR filter systems employed are small compared to the photometric zero
point errors for each cluster. We stack all cluster and cluster
canditates as follows: spectroscopically confirmed clusters at $z>0.8$
(5 systems), cluster candidates at $0.8<z_{phot}<1.2$ (4), cluster
candidates at $z_{phot}>1.2$ either with a clear red sequence (4) or
without (3, i.e. clusters 16, 17 and 18).  We also stack the CMD of
six control fields constructed in order to test the null hypothesis
that each cluster candidate is false. We take the location of six
clusters observed with HAWK-I and shift the cluster centroid to an
adjacent detector (approximately a 3\arcmin\ shift). We then repeat
the stacking procedure using these new centroids and apply the same
colour and magnitude shifts applied to galaxies selected according the
original cluster locations.

\begin{figure}
\centering
\psfig{figure=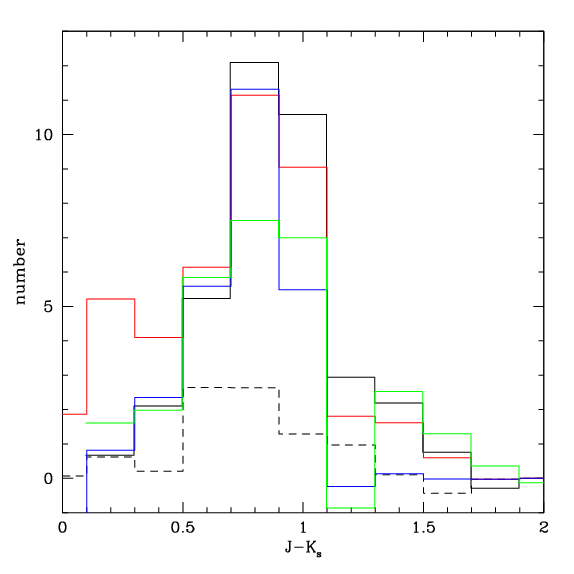,width=3.5in,angle=0.0}
\caption{Colour histograms generated by summing over the stacked CMD
  histograms for each cluster subsample: spectroscopically confirmed
  clusters (solid black; 5 systems), candidate clusters with
  $z_{phot}<1.2$ (green; 4 systems), candidate clusters at
  $z_{phot}>1.2$ with evidence for individual red sequences (red; 4
  systems) and without (blue; 3 systems), 6 random locations within
  the HAWK-I fields (dashed black). The numbers represent the average
  number of galaxies at each colour brighter than the applied
  brightness cut. See text for further details.}
\label{fig:chist}
\end{figure}

Figure \ref{fig:chist} displays the colour histrogram generated from
each stack by summing the CMD along the magnitude axis with the
restriction $K_s<22.5$, for $J-K_s<0.5$, or $K_s<23.0 - (J-K_s)$ otherwise, to
consider the photometrically complete region of the CMD. Figure
\ref{fig:chist} confirms that there is little difference in the
intrinsic red sequence properties between spectroscopically confirmed
clusters at $z\sim1$ and the candidate clusters. More importantly, the
subset of candidate clusters lacking apparent red sequences on the
basis of photometric redshift selection have been shown to have
average red sequences statistically identical to the remainder of the
candidate sample following the stacking analysis. We speculate that
these clusters display a range of star formation histories that are
not well described by the available SED templates used in the
photometric redshift analysis. The resulting photometric redshift peak
will be broadened by this systematic uncertainty and attempting to
select cluster galaxies using the photometric redshift method
described in this paper will result in a greater level of background
contamination relative to clusters with well modeled SEDs. Thus the
already weak red sequence may be diluted further by the increased
effective background along these sightlines. This effect can only be
verified once spectroscopic redshifts are available for these
clusters. All of the stacked cluster colour distributions are clearly
real and significant when compared to the null distribution. For the
clusters without an individual red sequence the fact that only 3
systems contribute to the average supports the assertion that the
majority (if not all) are real as otherwise the signal observed in the
average histogram would be very much diluted.  We therefore conclude
that each of the clusters presented in this sample is real in that the
extended X-ray source is associated with a galaxy population clustered
both spatially and in colour, whose spectra, photometric redshifts
and/or colours are consistent with $z>0.8$.

\subsection{The cluster red fraction}

The cluster red fraction provides a simple statement of the population
mix of cluster galaxy members selected by colour and magnitude.  It is
conceptually identical to the cluster blue fraction computed by
Butcher and Oemler (1984) yet here we focus on the the red galaxy
component in order to highlight the contribution of the cluster red
sequence to each cluster in our sample. We compute the red fraction as
the ratio
\begin{equation}
{
f_R = \frac{N_{R,cluster} - N_{R,back}}{N_{T,cluster}-N_{T,back}}
}
\end{equation}
where $N_R$ denotes the number of galaxies satisfying $J-K_s>\rm 3RG
-0.3$ and $K_s$ brighter than a evolving model early-type galaxy.  The
model assumes a galaxy of $K_s=22.5$ at $z=1.5$ described by a 1Gyr
solar metallicity burst of star formation occurring at $z_f=10$ and
described by a Salpeter IMF.  The corresponding $K_s$ limit at the
redshift of each cluster is computed accordingly.  $N_T$ denotes the
total number of galaxies satisfying the appropriate magnitude
limit. $N_{cluster}$ includes all galaxies within 1\arcmin\ of the
X-ray centroid and $N_{back}$ includes all galaxies at greater than
1\arcmin\ from the cluster centroid with the value scaled to match the
relative areas of the cluster and background samples.

\begin{figure}
\centering
\psfig{figure=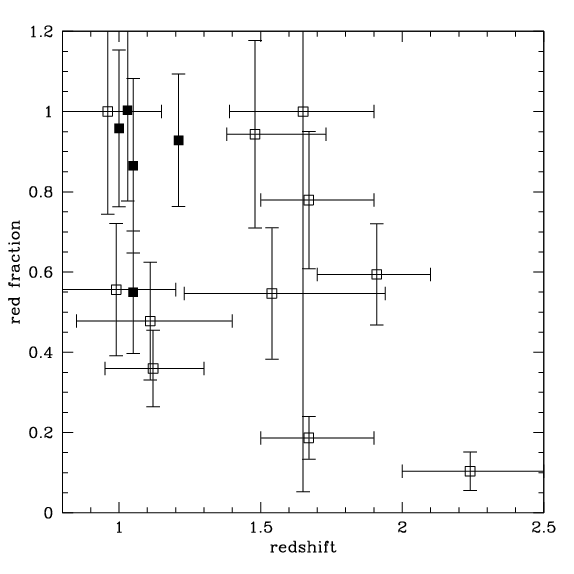,width=3.5in,angle=0.0}
\caption{The red fraction for all confirmed (solid squares) and
  candidate (open squares) $z>0.8$ clusters with $J-K_s$
  photometry. Error bars on the red fraction are Poissonian.}
\label{fig_red_frac}
\end{figure}

The red fraction values for all confirmed and candidate clusters with
$J-K_s$ photometry are displayed in Figure \ref{fig_red_frac}. The
figure indicates that the cluster sample displays a range of red
fraction values ranging from clusters almost wholly dominated by the
red sequence ($f_R\sim1$) to those with low values, i.e. $f_R<0.2$.
The observation of a wide range of red fraction values supports the
assertion that the compilation of a complete sample of X-ray selected
distant clusters can provide a relatively unbiased view of galaxy
populations in such systems. A comparable analysis of the red
fractions in a non-X-ray selected distant cluster sample has not yet
been performed \---\ which is unfortunate as the results of such a
study would provide a valuable perspective on the wavelength dependent
biases affecting distant cluster identification.

The range of red fraction values displayed by the distant cluster
sample is similar in extent to that of comparable mass clusters at
redshifts $z\sim0.3$ \citep{urq2010} also studied within the XMM-LSS
survey (note that we discuss mass estimation based upon X-ray flux
measurements in Section \ref{sec_xflux}).  However, the XMM-LSS
distant clusters will increase in mass by a factor $\sim5$ between a
redshift $z=1.5$ and $z<0.3$ \citep{boylan2009}.  Such low redshift
clusters of mass $> 5 \times 10^{14} \rm M_\odot$ typically display
dominant, bright red sequence populations with $f_R>0.8$.  If one
assumes that the evolution of galaxies onto the red sequence occurs
following the rapid cessation of star formation (quenching) \---\ for
example as a result of ram pressure stripping \citep{gunn1972} or
galaxy-galaxy interactions \citep{dressler1994} \---\ then the large
observed range of red fraction values displayed by the XMM-LSS distant
cluster is consistent with the scenario whereby the galaxy populations
have been caught in a variety of states transforming between active
star forming environments (where the red fraction is low and
comparable to that of the field) to a ``red and dead'' environment
typified by $z<1$ massive clusters.

\subsection{X-ray fluxes and cluster masses}
\label{sec_xflux}

A key motivation for identifying the most distant clusters is to
determine their global properties and thereby reveal the details of
cluster evolution. One of the most important properties is the total
cluster mass.  Unfortunately, this is not observable directly but it
can, with certain assumptions, be determined from other observables
including the X-ray flux (luminosity) of a system.  While one could
employ well-studied scaling relations at $z<1$ in order to convert a
flux measurement to one of mass, the extrapolation of these scaling
relations to $z>1$ is fraught with uncertainty. We therefore adopt an
alternative approach whereby we compare the X-ray flux observed from
clusters of known redshift (either spectroscopic or photometric) to
the flux expected from a model cluster of specified properites. This
approach offers the reassurance that the model assumptions are defined
and compared to the two observables (flux and redshift) in as clear a
manner as possible.

\begin{figure}
  \centering
  \psfig{figure=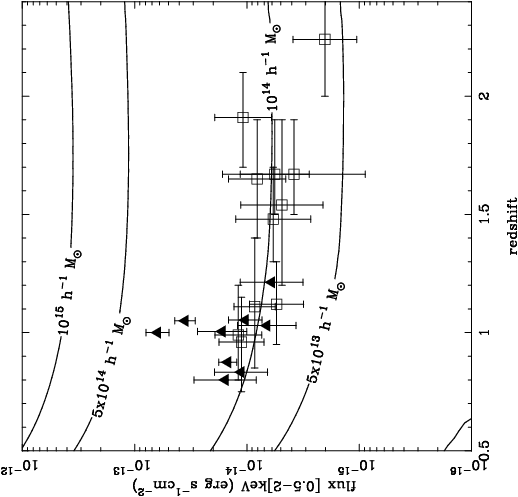,width=3.0in,angle=270.0}
  \caption{X-ray flux measurements for clusters with confirmed
    spectroscopic redshifts $z>0.8$ (filled triangles) and candidate
    clusters with photometric redshifts $z_{phot}>0.8$ (open
    squares). The contours indicate the expected flux versus redshift
    properties of model clusters of given mass (see text for
    details).}
\label{fig_xflux_redshift}
\end{figure}

Figure \ref{fig_xflux_redshift} compares the flux values for
individual clusters to the flux of model clusters computed as a
function of cluster mass and redshift.  The model assumes the
luminosity-temperature relation of \citet{arnaud99} with self-similar
evolution. The mass-temperature relation is taken from
\citet{arnaud05} with $\delta=200$ and massive clusters (extrapolated
down to the entire mass range) with self-similar evolution.  A
comparison of the flux values for the individual clusters to the model
indicates that the clusters display an approximate mass limit of $6
\times 10^{13}$ to $1 \times 10^{14} \rm M_\odot$. All clusters
display masses inferred from this comparison less than $4 \times
10^{14} \rm M_\odot$. We refer to this as the baseline model in the
following text.

\begin{figure}
  \centering
  \psfig{figure=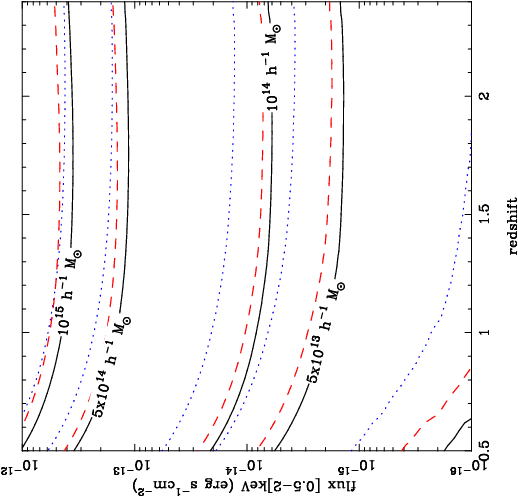,width=3.0in,angle=270.0}
  \caption{A comparison of the predicted flux versus redshift
    behaviour of model clusters of varying assumed scaling
    relations. Black solid lines indicate the baseline model of
    \citet{arnaud99} and \citet{arnaud05}. Blue dotted lines indicate
    a model assuming a $M_{500}-T$ relation from Sun et al. (2009) and
    a $L_{500}-T$ relation from Pratt et al. (2009). The red dashed
    lines indicate a model employing scaling laws described in
    Vikhlinin et al. (2009). See text for further details.}
\label{flux_redshift_compare_self}
\end{figure}

\begin{figure}
  \centering
  \psfig{figure=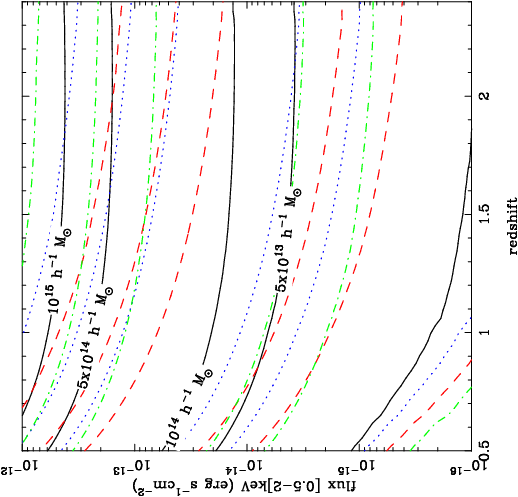,width=3.0in,angle=270.0}
  \caption{A comparison of the predicted flux versus redshift
    behaviour of model clusters of varying assumed evolutionary
    properties. The black solid lines indicate the baseline model
    described in the text which assumes self-similar
    evolution. Alternative evolution prescriptions include that of
    Reichert et al. (2011; blue dots), Clerc et al. (2012; red dashed)
    and the evolving scaling relations derived by Vikhlinin et
    al. (2009; green dot-dash).}
\label{flux_redshift_compare_evol}
\end{figure}

It is clear that adopting a different set of model assumptions will
affect the mass estimates returned by the above analysis.  Figure
\ref{flux_redshift_compare_self} examines the extent to which adopting
an alternative set of scaling relations influences the estimated
cluster mass.  We retain the assumption that scaling laws evolve in a
self-similar manner and compare two alternative approaches to our
baseline model described above (Figure
\ref{flux_redshift_compare_self}; black solid lines): 1) a model which
replaces the $M_{200}-T$ relation for a $M_{500}-T$ relation taken
from Sun et al. (2009), valid down to 1 keV, and assuming the
$L_{500}-T$ relation described by Pratt et al. (2009), see Figure
\ref{flux_redshift_compare_self} (blue dotted lines).  Model 2)
considers the flux-redshift relation for self-similar clusters
following the scaling laws described in Vikhlinin et al. (2009)
\---\ see Figure \ref{flux_redshift_compare_self} (red dashed lines).
Each of these alternative models generates mass estimates for clusters
of given flux and redshift which are a generally a factor 2 lower than
those generated applying the baseline model.
Figure \ref{flux_redshift_compare_evol} examines the extent to which
allowing a given scaling relation to evolve with redshift affects the
estimated cluster mass.  Once again, the black lines in Figure
\ref{flux_redshift_compare_evol} indicate the baseline model described
by self-similar evolution.  Alternative evolution prescriptions
include that of Reichert et al. (2011; blue dots), Clerc et al. (2012;
red dashed) and the evolving scaling relations derived by Vikhlinin et
al. (2009; green dot-dash).  In this case, introducing alternative
assumptions regarding the evolution of scaling relations can generate
mass estimates for clusters of given flux and redshift which are up to
a factor 2-4 greater than those generated applying the baseline model.
The above considerations indicate that models which consider potential
variations in cluster scaling relations and their evolution can
generate mass estimates for $z>0.8$ clusters which vary by up to an
order of magnitude.  However, given the shape of the contours
displayed in Figures \ref{fig_xflux_redshift},
\ref{flux_redshift_compare_self}, and \ref{flux_redshift_compare_evol}
it is apparent that the {\it relative} masses of the $z>0.8$ clusters
are likely to be robust against uncertainties in the assumed cluster
scaling relation model.

\subsection{The abundance of distant X-ray selected clusters}

Having compiled a sample of $z>0.8$ confirmed and candidate clusters
it is instructive to compare their observed abundance to the number
predicted using a calculation based upon all relevant cosmological,
cluster scaling relation and selection considerations.  We computed
the expected abundance of distant clusters by assuming a
\citet{Tinker2008} redshift-dependent mass function. These authors
provide a functional form of the halo mass distribution calibrated
using numerical simulations up to $z\sim 2.5$. For this calculation,
we assume $\sigma_8 =0.787$ \citep{dunkley09}. Halo masses are then
converted from $M_{200b}$ to $M_{200c}$ following
\citet{hu_kravtsov_2003} assuming a NFW mass profile and a
concentration model from \citet{bullock2001}.  The conversion from
masses to X-ray observables is performed based upon the same baseline
$M_{200c}-T_X$ and $L_X-T_X$ scaling laws as quoted above. We assume a
scatter of $\sigma_{\ln L_X | T_X} = 0.6$ in the $L_X-T_X$
relation. Both scaling laws are assumed to evolve self-similarly.  The
XMM-LSS C1+C2 selection function \citep{pacaud06} accounts for the
probability of detecting a cluster given its X-ray observables, namely
its [0.5-2] keV count-rate and its apparent core radius (corresponding
to a $\beta$ model with $\beta=2/3$). For this purpose, model
count-rates are estimated assuming an APEC spectral model with
abundance $Z=0.3 Z_{\odot}$. The core radius is assumed to scale with
the halo radius following a simple scaling relation
\citep{clerc2011b}: $R_c = 0.24 \times R_{500}$.

The final expected redshift distribution is integrated in various
redshift bins: we predict 28.5 clusters lying at $z>0.8$, 5.2 at
$z>1.3$ and 2.5 at $z>1.5$ in 9 deg$^2$ of surveyed area. By
comparison, the sample of high redshift clusters presented in this
work consists of up to 22, 7 and 6 clusters at $z> 0.8$, $1.3$ and
$1.5$ respectively, thus showing a rough agreement with model
predictions. However we note a slight ($\sim 2 \sigma$) excess of
$z>1.5$ clusters in our sample. If confirmed, this excess could be due
either to an overestimate of photometric redshifts or to a unaccounted
selection bias that would arise due to e.g. an increased contamination
of the X-ray flux from unresolved AGNs in these objects.  Furthermore,
model uncertainties also impact the predicted number of high-redshift
clusters.  In particular, relaxing the self-similarity constraint on
the evolution of the cluster mass-luminosity relation can lead to
considerably different predictions \citep{pacaud07}.

A further comparison of interest is that between the surface density
of XMM-LSS distant clusters and that of other X-ray selected distant
cluster samples present in the literature.  The surface density of
$z>0.8$ clusters in the XMM-LSS survey is $2.3\pm0.5\,\rm deg^{-2}$
above a nominal mass limit of $\sim1 \times 10^{14} \rm M_\odot$.
This may be compared to a figure of 15 $z>1.1$ clusters in
approximately 1 \dd reported by \citet{bielby10} above a mass limit of
$\sim1 \times 10^{14} \rm M_\odot$ In addition \citet{fass11} present
a compilation of 22 $0.9<z<1.6$ clusters detected in up to 79 \dd of
archival XMM observations with a mass limit of $1-2 \times 10^{14} \rm
M_\odot$ (though specific details regarding the selection function are
currently unavailable). The surface density of clusters reported in
\citet{fass11} lies between 0.3 and 1.3 $\rm deg^{-2}$ with the exact
value depending upon the subset of XMM observations analysed. We note
that the \citet{fass11} compilation containes two previoulsy published
XMM-LSS clusters with the result that this comparison is largely but
not completely independent.

The variance observed in these reported surface densities arises from
differences in the techniques applied to select extended X-ray
sources, confirm galaxy overdensities and subsequently compute
photometric redshifts or to compile spectroscopic redshifts.  A
further point worth noting is that distant clusters are often detected
in survey data originally compiled to study galaxy clusters at
$z<1$. As such, they represent a subset of marginal detections and
whether they are subsequently classified and confirmed as distant
clusters is a very sensitive function of the set of selection tests
applied to the data.  With this in mind we have attempted to generate
a complete sample of distant X-ray clusters in a manner that depends
solely upon the X-ray data by performing an analysis of all extended
sources in a subset of the XMM-LSS area.  One cannot completely
escape the requirement of input from other wavebands, e.g. the
optical, NIR and MIR data employed in this paper. However, we have
attempted as far as possible to select galaxy overdensities associated
with each extended X-ray source in a manner which is insensitive to
the assumed star formation history of individual galaxies.

More detailed follow-up of individual clusters, including
spectroscopic and deeper X-ray imaging observations are currently
underway. Ultimately combining the precise redshift measurements into
a self-consistent analysis including selection effects and model
uncertainties \citep{pacaud07, maughan2012, reichert2011, clerc2011b},
we will be able to draw firm conclusions from the abundance of
high-redshift clusters.

\section{Conclusions}
\label{conclude}

The analysis presented in this paper effectively completes the
assessment of 88 extended C1 and C2 sources from approximately 9 \dd
of XMM-LSS data.  Of these sources 59 display spectroscopic or
photometric redshifts $z<0.8$, 21 sources display spectroscopic or
photometric redshifts $z>0.8$ (or colours consistent with the same
redshift limit in the case of candidate clusters 23 and 24), and a
remaining 8 sources appear to be consistent with misclassified point
sources or marginal detections.  The sample also contains cluster 19
at a photometric redshift $z=1.67$ which is included in this paper
having been flagged as a potential distant cluster at an earlier
stage.  The distant cluster sample is complete in that it represents
(with the low-redshift and marginal source identifications) a complete
account of a 9 \dd area of the XMM-LSS survey.  This sample is
generated from the X-ray data employing a quantitative selection
function \citep{pacaud06} and it therefore permits a number of
important questions in cosmology and galaxy evolution to be
investigated.

The complete nature of this sample is dependent upon the primacy of
the applied X-ray selection procedures. Although it is difficult to
conceive of a targeted X-ray survey for distant clusters that does not
employ information at additional wavebands (e.g. optical, MIR, etc.),
the application of a simple selection threshold to identify
high-significance clusters in these additional wavebands (e.g. the
surface density of optical-MIR selected sources) will either lead to
an incomplete X-ray sample or a large rate of contamination (from
low-redshift or spurious sources), c.f. Figure \ref{fig_apc}.  These
comments do not undermine the nature of optical-MIR selected samples
of distant clusters \---\ which are complete in terms of the applied
selection criteria. Instead they reflect the fact that distant X-ray
detected clusters must be confirmed at other wavebands.  In doing so
with this paper we have attempted to perform as comprehensive an
assessment as possible of each confirmed and candidate system with the
aim of compiling a complete sample of X-ray selected distant clusters.

It is important to recognise that the analysis presented in this paper
represents only one stage in the creation a complete sample of distant
X-ray clusters. Clearly, much of the interpretation as to the nature
of each cluster rests upon the photometric redshift analysis and
spectroscopic confirmation of the redshifts of these clusters must be
considered as the next, important step.  Furthermore, the extent to
which point source emission from unresolved AGN (both within each
cluster and superposed along the line of sight) modifies the
appearance of a sample of distant X-ray clusters is not well
understood. We intend to employ the superior angular resolution of the
Chandra observatory to characterise the point source contribution to a
representative sub-sample of the distant clusters presented in this
paper.  Only when these steps are complete will we have a better
understanding of what constitutes a complete, robust sample of distant
clusters.
This XMM-LSS distant cluster sample represents an important resource
and will form the basis for studies in both the growth of large scale
structure and the evolution of cluster galaxies to be presented in
forthcoming papers.

\section*{Acknowledgments}

The authors wish to thank Joana Santos, Chris Lidman, Graham Smith and
Emanuele Daddi for useful discussions during the development of this
paper. JPW acknowledges financial support from the Canadian National
Science and Engineering Research Council (NSERC).

This work is based on observations obtained with XMM-Newton, an ESA
science mission with instruments and contributions directly funded by
ESA Member States and the USA (NASA). Based on observations collected
at the European Organisation for Astronomical Research in the Southern
Hemisphere, Chile (program IDs 72.A-0104, 84.A-0740 and
86.A-0432). Based on observations obtained at the Gemini Observatory
(Program ID GS-2006B-Q-22), which is operated by the Association of
Universities for Research in Astronomy, Inc., under a cooperative
agreement with the NSF on behalf of the Gemini partnership: the
National Science Foundation (United States), the Science and
Technology Facilities Council (United Kingdom), the National Research
Council (Canada), CONICYT (Chile), the Australian Research Council
(Australia), Ministério da Ciência, Tecnologia e Inovação (Brazil) and
Ministerio de Ciencia, Tecnología e Innovación Productiva
(Argentina). Based on observations obtained with WIRCam, a joint
project of CFHT, Taiwan, Korea, Canada, France, and the
Canada-France-Hawaii Telescope (CFHT) which is operated by the
National Research Council (NRC) of Canada, the Institute National des
Sciences de l'Univers of the Centre National de la Recherche
Scientifique of France, and the University of Hawaii. Based on
observations obtained with MegaPrime/MegaCam, a joint project of CFHT
and CEA/DAPNIA, at the Canada-France-Hawaii Telescope (CFHT) which is
operated by the National Research Council (NRC) of Canada, the
Institut National des Sciences de l'Univers of the Centre National de
la Recherche Scientifique (CNRS) of France, and the University of
Hawaii. This work is based in part on data products produced at
TERAPIX and the Canadian Astronomy Data Centre as part of the
Canada-France-Hawaii Telescope Legacy Survey, a collaborative project
of NRC and CNRS.

\bsp

\label{lastpage}

\end{document}